\documentclass[aps,prd,nofootinbib,twocolumn,superscriptaddress,preprintnumbers,balancelastpage,longbibliography]{revtex4-2}

\usepackage{multirow}
\usepackage{float}
\usepackage{graphicx} 
\usepackage{url}
\usepackage{graphicx}
\usepackage{amsmath,amssymb,amstext,amssymb,amsfonts,amsthm}
\usepackage{hyperref}
\usepackage{appendix}
\usepackage{physics}
\usepackage{soul}
\usepackage{comment}
\usepackage{color}
\usepackage[normalem]{ulem}
\usepackage{lipsum}
\usepackage{capt-of}
\usepackage{aas_macros}

\usepackage{todonotes}

\renewcommand{\dd}{\mathrm{d}}
\newcommand{\vect}[1]{\boldsymbol{\mathbf{#1}}}
\definecolor{deepgreen}{rgb}{0.2,0.8,0.2}

\definecolor{deepblue}{rgb}{0.2,0.2,0.8}

\definecolor{deepred}{rgb}{0.8,0.2,0.2}

\begin{document}
\title{Expanding Ejecta Method:\\
II.~Framework for Cosmological Distance Measurements via Intensity Interferometry}
\author{David Dunsky}
\email{ddunsky@nyu.edu}
\affiliation{Center for Cosmology and Particle Physics, Department of Physics, New York University, New York, NY 10003, USA}

\author{I-Kai Chen}
\email{ic2127@nyu.edu}
\affiliation{Center for Cosmology and Particle Physics, Department of Physics, New York University, New York, NY 10003, USA}
 
\author{Junwu Huang}
\email{jhuang@perimeterinstitute.ca}
\affiliation{Perimeter Institute for Theoretical Physics, 31 Caroline St.~N., Waterloo, Ontario N2L 2Y5, Canada}

\author{Ken Van Tilburg}
\email{kenvt@nyu.edu, kvantilburg@flatironinstitute.org}
\affiliation{Center for Cosmology and Particle Physics, Department of Physics, New York University, New York, NY 10003, USA}
\affiliation{Center for Computational Astrophysics, Flatiron Institute, New York, NY 10010, USA}

\author{Robert V.~Wagoner}
 \email{wagoner@stanford.edu}
 \affiliation{Department of Physics and KIPAC, Stanford University, Stanford, CA 94305, USA}

\date{\today}

\begin{abstract}
We explore the potential of the expanding ejecta method (EEM)~\cite{eem-1} as a cosmological probe, leveraging its ability to measure angular diameter distances to supernovae (SNe) with intensity interferometry. We propose three distinct applications of the EEM: 
(1) using Type IIP SNe as moderate-distance geometric anchors to calibrate Cepheids, replacing other local distance indicators; 
(2) directly calibrating Type Ia SNe, bypassing conventional calibration methods;
(3) constructing a fully independent Hubble diagram with Type IIP (Type Ia) SNe, entirely decoupled from the traditional distance ladder.
Incorporating realistic SN populations, we forecast a Hubble constant precision with next-generation intensity interferometers of $1.6\%$, $1.1\%$, and $9.3\% \,(3.6\%)$, respectively, for the three different proposed applications. Future intensity interferometry could yield improvements to $1.2\%$,  $0.6\%$, and $1.5\%\,(0.4\%)$. The EEM thus offers a powerful geometric alternative for cosmic distance determination.
\end{abstract}
\maketitle

The hypothesis of cold dark matter (CDM) and a time-independent dark energy pervading the Universe alongside known matter has been spectacularly successful in explaining observations and is widely accepted as the standard model of cosmology. 
The simplest underlying microphysical theory for $\Lambda$CDM is that of a gravitationally-interacting massive particle and a cosmological constant ($\Lambda$), with no conclusive evidence for observable deviations from $\Lambda$CDM emerging to date, notwithstanding tentative anomalies \cite{2023OJAp....6E..36D, 2022PhRvD.105d3517P, 2024arXiv241112022D, 2025arXiv250314738D}.

One of the fundamental parameters of $\Lambda$CDM is the total energy density of the Universe, or equivalently, the present-day cosmic expansion rate $H_0$. 
The cosmic microwave background (CMB) provides a model-dependent determination of $H_0$ by constraining the angular sound horizon $\theta_*$ of the baryon acoustic oscillation peak at recombination~\cite{1996PhRvD..54.1332J}:
$\theta_* = {r_s(z_*)}/{D_A(z_*)}$,
where $r_s(z_*) $ is the sound horizon at last scattering (redshift $z_*$), and $D_A(z_*)$ the angular diameter distance to the CMB. Within $\Lambda$CDM, $ r_s(z_*)$ is set by early-Universe physics, while $D_A(z_*)$ depends on the entire expansion history, effectively making $H_0$ an extrapolated late-time parameter~\cite{Aylor:2018drw}. 
In contrast, $H_0$ can be directly determined in the late Universe by measuring the luminosity distance $D_L = (1+z) c \int_0^z \, \mathrm{d} z' / {H(z')}$ and redshift $z$ of Type Ia supernovae (SNe Ia), which serve as standardizable candles, allowing an inference of the redshift-dependent Hubble parameter $H(z) = H_0 \sqrt{\Omega_m (1+z)^3 + \Omega_\Lambda}$ in a flat $\Lambda$CDM universe. This procedure requires an absolute calibration of the intrinsic luminosity of SNe Ia, typically tied to local distance indicators such as Cepheid variables~\cite{1991PASP..103..933M} or the tip of the red giant branch (TRGB)~\cite{2007ApJ...661..815R}, which are themselves anchored to local distance indicators such as parallax~\cite{2018ApJ...861..126R,2021ApJ...908L...6R}, eclipsing binaries~\cite{2019ApJ...876...85R}, and masers~\cite{2020ApJ...891L...1P}. The combination of these three rungs makes up the cosmic distance ladder (CDL).

Improvements in purported precision have exposed a ``Hubble tension'', with different methodologies yielding $H_0$ estimates with a fractional variation of around 7\% (see reviews~\cite{2010ARA&A..48..673F,2023ARNPS..73..153K}). The persistent discrepancy between CMB results~\cite{2013ApJS..208...20B, 2020A&A...641A...6P,2021PhRvD.104b2003D, 2020JCAP...12..047A}, $H_0 = 68.22 \pm 0.36\,\mathrm{km/s/Mpc}$~\cite{2025arXiv250314452L}, and many (but not all) local SN-based determinations~\cite{2024arXiv240806153F, 2022ApJ...934L...7R}, $H_0 = 73.17\pm 0.86\,\mathrm{km/s/Mpc}$~\cite{2024ApJ...973...30B}, has prompted the search for alternative, independent methods to cross-check and calibrate cosmic distance measurements. Approaches such as gravitational wave standard sirens \cite{1986Natur.323..310S,Oguri:2016dgk,2018Natur.562..545C}, surface brightness fluctuations~\cite{1988AJ.....96..807T}, Mira variables~\cite{2020ApJ...889....5H}, strong-lensing time delays~\cite{1964MNRAS.128..307R,2020A&A...643A.165B}, fast radio burst interferometry~\cite{2023ApJ...947L..23B}, and intensity interferometry measurements of active galactic nuclei~\cite{Dalal:2024aaj} each offer unique systematics and complementary insights.  

One such alternative for the local distance ladder, the expanding photosphere method (EPM) \cite{KirshnerKwan,1992ApJ...395..366S,1973ApJ...185..303K,1994ApJ...432...42S}, estimates SN distances by assuming a diluted blackbody photosphere and leveraging spectral information to infer its expansion properties. However, the EPM is limited by uncertainties in asphericity, extinction, radiative transfer, and unknown deviations from a perfect blackbody, leading to systematic errors that have prevented it from becoming a primary cosmological tool~\cite{1981ApJ...250L..65W,1993PhR...227..205W,2017hsn..book..769S,filippenko1997optical,2023ApJ...942...38M,Bartel:2007px}, although variants mitigate some of these effects~\cite{1995ApJ...441..170B,2004ApJ...616L..91B, 2002ApJ...566L..63H,2014AJ....148..107R,2015ApJ...815..121D,2023A&A...678A..14S, 2024arXiv241104968V}.

We propose an alternative: the \textit{expanding ejecta method (EEM)}, which circumvents key limitations of the EPM by directly resolving the expansion of SN ejecta using intensity interferometry. The EEM exploits the extraordinary angular resolution of intensity interferometers to measure the \emph{angular} expansion velocities $\dot{\theta}_\mathrm{ej}$ of the ejecta, in addition to their \emph{physical} expansion velocities $v_\mathrm{ej}$. Together, these yield a geometric determination of the angular diameter distance,
\begin{equation}
    D_A \simeq \frac{v_\mathrm{ej}}{\dot{\theta}_\mathrm{ej}}, \label{eq:D_A}
\end{equation}
independent of luminosity-based calibrations. Eq.~\eqref{eq:D_A} is a simplistic interpretation of the full set of observables that constrain both the angular morphology and the velocity structure of the SN ejecta, described in more detail in our companion paper~\cite{eem-1}. The combination of the integrated spectrum and intensity correlations provide information on the explosion properties. However, only intensity interferometry can break the degeneracy between physical and angular size to infer $D_A$, without resorting to assumptions about surface brightness, extinction, or spherical symmetry that have hampered the traditional EPM and its variants.

As a late-Universe method with uncorrelated uncertainties to the traditional rungs of the CDL, the EEM offers a promising avenue for precision cosmology, complementary to other approaches. 
In this work, we explore three distinct applications of the EEM: (1) calibration of Cepheids using Type IIP SNe as geometric distance anchors; (2) direct calibration of Type Ia SNe, circumventing conventional methods based on Cepheids or TRGB; and (3) construction of a completely independent Hubble diagram.
We assess the achievable precision for each of these three methods under realistic observational constraints, demonstrating that the EEM could serve as a powerful new tool for determining $H_0$ and mitigating systematics in cosmological distance measurements.

\paragraph*{\bf SN distance determination.---}
We start with a summary of the EEM~\cite{eem-1} and its observables, and quantify the precision to which $D_A$ can be inferred as a function of a SN's apparent magnitude. We will use these results later on for our forecasts on $H_0$ determinations.

The cataclysmic termination of a massive star or white dwarf as a SN expels solar masses of material away from the progenitor's original location, at a range of speeds of order $10^4\,\mathrm{km/s}$.
The radiative transfer modeling of how electromagnetic radiation, produced by shock heating and radioactive decay, escapes the complex 3D distribution of the expanding ejecta is a subject of intense investigation~\cite{1996ApJ...466..911E, 2005A&A...439..671D, 2011MNRAS.415.3497D, 2015MNRAS.453.2189D, 2024A&A...684A..16D}. However, the basic structure is that of a (potentially wavelength-dependent) photosphere surrounded by ballistically expanding ejecta.

The characteristic radius of a SN IIP's photosphere is roughly $R_\mathrm{ph} \sim 10^{15}\,\mathrm{cm}$ in the plateau phase, corresponding to an angular radius:
\begin{align}
    \Theta_\mathrm{ph} = \frac{R_\mathrm{ph}}{D_A} \approx \underbrace{10^{-11}\,\mathrm{rad}}_{2\,\mathrm{\mu as}} \left( \frac{30 \, \mathrm{Mpc}}{D_A} \right) \left( \frac{R_\mathrm{ph}}{10^{15}\,\mathrm{cm}} \right).
\end{align}
Spatially resolving such a tiny emission region is beyond the capabilities of imaging telescopes and amplitude interferometers, but would be an ideal target for an intensity interferometer (array) with a characteristic baseline $d \simeq 20\,\mathrm{km}$, yielding a fiducial resolution of $\lambda / d \approx 3\times 10^{-11}\,\mathrm{rad}$ on the $\mathrm{H\alpha}$ line with wavelength $\lambda = 656\,\mathrm{nm}$ at the optimal operating point~\cite{eem-1}.

\begin{figure}
\includegraphics[width=.5\textwidth]{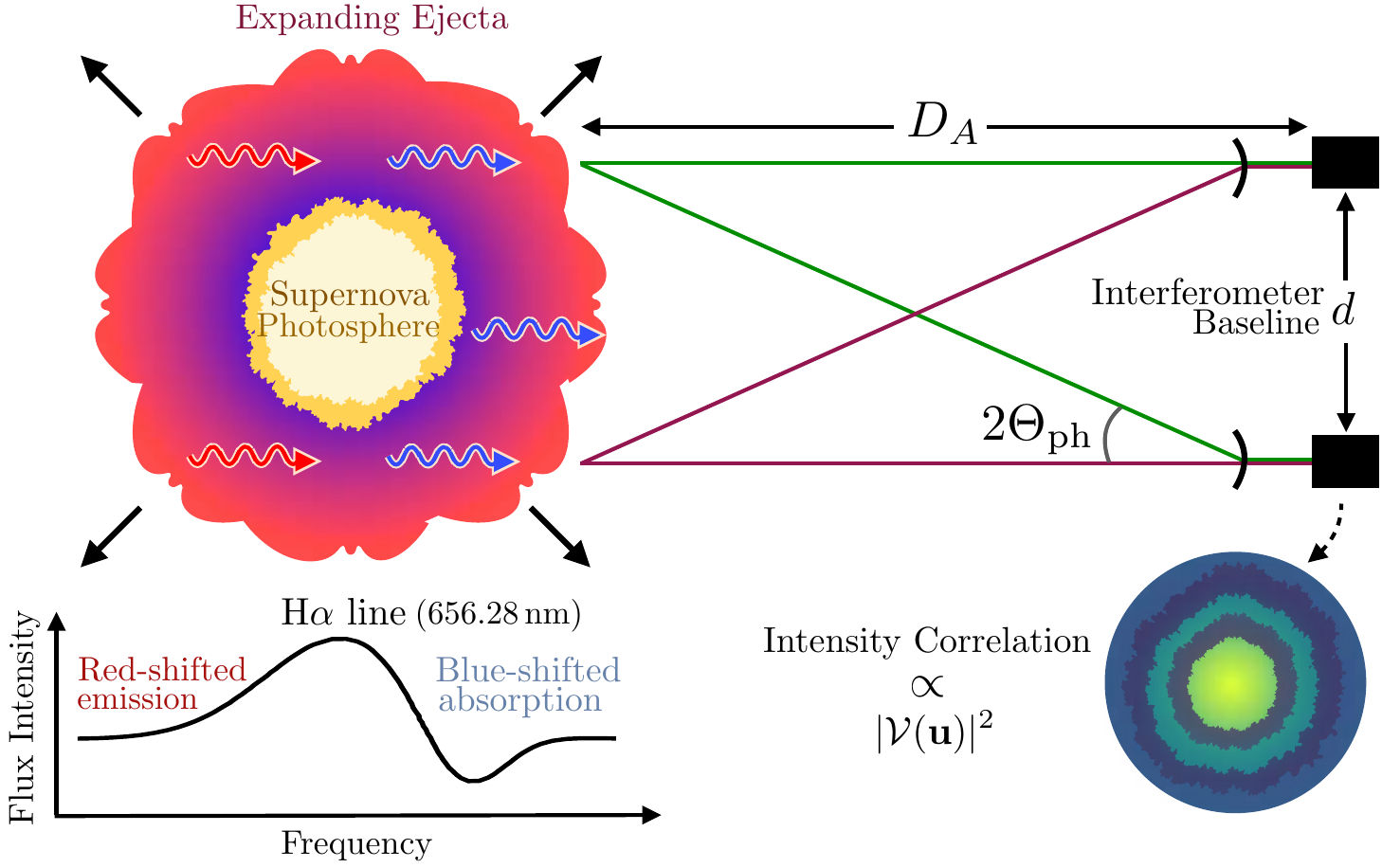}
\caption{
EEM illustration. The SN photosphere (yellow) and ejecta (blue to red) are resolved by a spectrally-multiplexed intensity interferometer with a large baseline $d$, in a wavelength band around a spectral line. The ejecta are illuminated by the photosphere and produce a P-Cygni profile in the flux density (left inset). The angular information in the square modulus of the visibility function $\mathcal{V}$ as measured from the correlated intensities (right inset), combined with the velocity information in the spectrum,  enables inference of the SN morphology and angular diameter distance $D_A$.
}
\label{fig:eemDiagram}
\end{figure}

The key idea is that resolved spatial information over a set of narrow spectral channels---the square moduli of the visibility function---constrains the unknown shape of the SN photosphere and the distribution of the ejecta. By fitting a parametric model of this morphology and radiative transfer in both angular and velocity space, $D_A$ can be inferred through Eq.~\eqref{eq:D_A}. The model can be improved self-consistently and is robust to uncertainties in luminosity calibration, flux dilution, or extinction in the host environment. With sufficient signal-to-noise ratio (SNR) on the visibility function moduli of a bright nearby SN, model-independent phase and thus image reconstruction may also be possible~\cite{2015A&A...580A..99D,IIAI}.

To understand the method qualitatively, consider a single parcel $i$ of ejecta material with a strong spectral line at rest wavelength $\lambda_0$ (e.g.~$\mathrm{H\alpha}$) moving away from the explosion center ballistically, illuminated by the (broadband) emission of the photosphere.  This produces a blue-shifted absorption line or a blue- or red-shifted emission line at the angular position $\vect{\theta}^i(t)$ of the parcel, and at a fixed shifted wavelength of $\lambda^i = \lambda_0 ( 1 + v^i_\parallel)$, where $v^i_\parallel$ is the parcel's line-of-sight velocity. The sum of all ejecta parcels illuminated by the photosphere results in the spectral intensity $I_\lambda(\lambda, \vect{\theta})$, about which \emph{all} information is contained in the integrated flux density $f(\lambda) = \int \dd^2 \vect{\theta}\, I_\lambda$ and the visibility function $\mathcal{V}(\lambda, \vect{u}) = f^{-1} \int \dd^2 \vect{\theta} \, e^{i \vect{u} \cdot \vect{\theta}} I_\lambda$, with $\vect{u} = 2\pi \vect{d}_\perp / \lambda$~\cite{eem-1, 2025arXiv250307725K}. We illustrate these concepts in Fig.~\ref{fig:eemDiagram}.

In ref.~\cite{eem-1}, we forecast the precision to which intensity interferometry can determine the parameters of a model with an ellipsoidal photosphere and arbitrary orientation, surrounded by a distribution of ejecta whose opacity falls off as a power law with arbitrary amplitude and steep spectral index, in a 10\% fractional bandwidth around a single spectral line (H$\alpha$ line for Type IIP SNe). Our Fisher analysis of SNe of varying sizes, asphericities, and orientations demonstrates that that there is sufficient information to break all degeneracies, in spite of sparse coverage in the $\vect{u}$ plane, lack of visibility phase information in intensity correlation, and the 3D explosion being viewed in 2D projection. In particular, the potentially worrisome ambiguity between a LOS dilatation (which transforms $v_\parallel$ but not $\dot{\vect{\theta}}$) and an \emph{intrinsic} asphericity in the LOS direction can be resolved. The EEM can provide precise determinations of both a SN's shape and angular diameter distance~\cite{eem-1}.
We project that an intensity interferometer array with two baselines, per-site light-collection area $A$, photodetector timing resolution $\sigma_t$, spectral channel resolution $\mathcal{R}$, and efficiency $\epsilon$, can measure a SN's angular diameter distance to a precision
\begin{align}
    \label{eq:sigmaD}
    \frac{\sigma_{D_A}}{D_A} &\approx 2 \% \times 10^{0.4(m-12)} \left(\frac{t_{\rm obs}}{60 \, \rm hr}\right)^{ \scalebox{.89}{-$\frac{1}{2}$}}   
     \\
    &\phantom{\approx} \times
    \left(\frac{\sigma_t}{10 \, \rm ps}\right)^{\scalebox{.89}{$\frac{1}{2}$}} \left(\frac{A}{\pi (5 \,{\rm m})^2}\right)^{-1}\left(\frac{\mathcal{R}}{10^4}\right)^{\scalebox{.89}{-$\frac{1}{2}$}}  \left(\frac{\epsilon}{0.5}\right)^{-1} \, , \nonumber
\end{align}
when combined with a $10\%$-precision spectral measurement of the flux density~\cite{eem-1}. 
This result is obtained with a concurrent marginalization over the morphology of the SN, which can be measured to a fractional precision of $7\%$ under the same assumptions. Following recent promising results~\cite{2025arXiv250307725K}, we (optimistically) assume that the same fractional uncertainty (at fixed apparent magnitude) can be achieved on SNe Ia in our subsequent projections.
For faint SNe, it may be useful to apply a prior on the asphericity gleaned from a sample of bright, nearby SNe, which can potentially improve distance uncertainties by a factor of up to 2. A weak but quantifiable selection bias on $D_A$ could exist when applying the EEM to a magnitude-limited sample of SNe (see App.~\ref{sec:dilation}), in particular when a prior on asphericity is introduced.
The first line of Eq.~\eqref{eq:sigmaD} indicates how  $\sigma_{D_A}/D_A$ scales with observational parameters: the apparent magnitude $m$ of the SN and the observation time $t_{\rm obs}$. The second line shows the scaling with the experimental parameters of the intensity interferometer array. For future convenience, we define its inverse according to:
\begin{equation}
    \text{Matchlight} \equiv 
    \left(\frac{\sigma_t}{1 \, \rm ps}\right)^{\scalebox{.825}{$-\frac{1}{2}$}} \left(\frac{A}{1 \,{\rm m}^2}\right)\left(\frac{\mathcal{R}}{1}\right)^{\scalebox{.825}{$\frac{1}{2}$}}  
    \left(\frac{\epsilon}{1}\right) \, , \label{eq:matchlight}
\end{equation}
which acts as a figure of merit for the capabilities of a given intensity interferometer array.

\begin{figure*}[t]
    \centering
     \includegraphics[width=2.075\columnwidth]{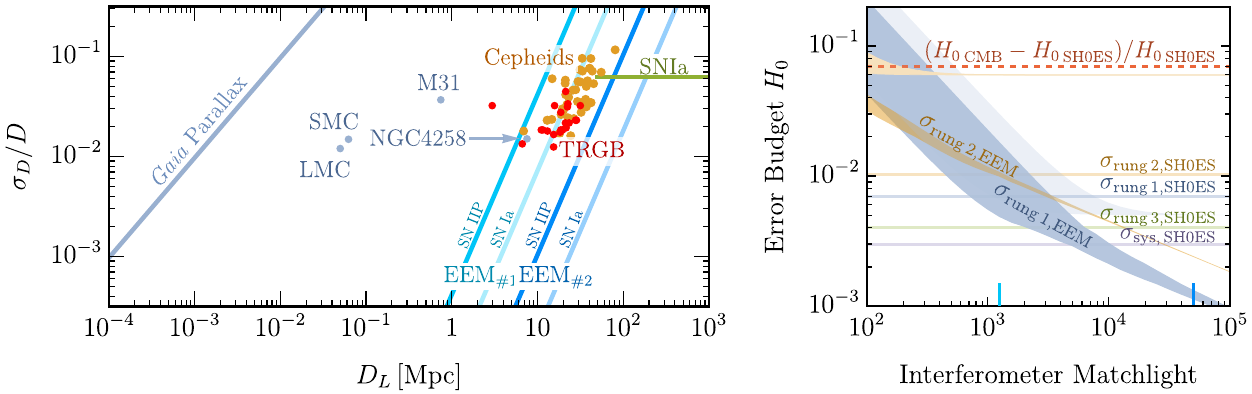}
\caption{\textit{Left:} Fractional distance precision $\sigma_D/D$ as a function of luminosity distance $D_L$. The light/dark blue lines are the EEM \#1/2 forecasts on SNe IIP for $\mathrm{Matchlight} = 1{,}250/50{,}000$ (blue ticks on right panel), respectively; lower-opacity lines are analogous SNe Ia extrapolations.
\textit{Gaia} parallaxes on \emph{individual} Cepheids can be statistically averaged to yield geometric distances to the Magellanic Clouds (blue dots); geometric anchors based on eclipsing binaries (M31) and masers (NGC4258) are also shown in blue. Distances based on the period-luminosity relation of Cepheids and TRGB are indicated as orange and red dots. Standard-candle distance precision to individual SNe Ia is shown as the green line.
\textit{Right:} Fractional precision contributions from the three rungs of the cosmic distance ladder (from SH0ES~\cite{2022ApJ...934L...7R}) in terms of $H_0$. The dark (light) blue band indicates the EEM-based calibration precision of Cepheids using all (only the brightest) SNe IIP over a 5-yr campaign. The dark (light) orange bands indicate the EEM calibration forecast of all (only the brightest) SNe Ia over 5~yr. The dashed red line signifies the current Hubble tension~\cite{2025arXiv250314452L,2024ApJ...973...30B}.
}
\label{fig:systemsDistanceUncertaintyPlot}
\end{figure*}

Since SNe of Type II (Type Ia) are roughly standard candles of absolute magnitude $M_{\rm SN} \approx -16.1 \, (-18.5)$~\cite{li2011nearby,Anderson:2014hta}, the dependence of $\sigma_{D_A}/D_A$ on apparent magnitude can be recast into a dependence on the luminosity distance $D_L = (1+z)^2 D_A \simeq 10^{(m - M_{\rm SN} -25)/5}~\mathrm{Mpc}$. Intrinsic variations in the SN absolute magnitude $M_{\rm SN}$ can affect the optimization of the SN population to target (as we elaborate in App.~\ref{sec:optimumSearchStrategyH0}), but they do not introduce a systematic bias in our measurement~\cite{1922MeLuF.100....1M}. The blue diagonal lines labeled ``EEM'' in Fig.~\ref{fig:systemsDistanceUncertaintyPlot} indicate the fractional distance precision as a function of $D_L$ for EEM application to Type II and Type Ia SNe for $t_{\rm obs} = 30 \times 6 \, \mathrm{hours}$. The light blue contours correspond to an interferometer array with $\mathrm{Matchlight} = 1{,}250$ $(\rm{EEM}_{\#1})$ as in  Eq.~\eqref{eq:sigmaD}, while the darker blue lines correspond to $\mathrm{Matchlight} =  50{,}000$~$(\rm{EEM}_{\#2})$.

Figure~\ref{fig:systemsDistanceUncertaintyPlot} also depicts the fractional precision of other methods that make up the CDL, including \textit{Gaia}'s parallax uncertainty on individual Cepheids~\cite{esaGaiaMission,2021ApJ...908L...6R,Reyes:2022boz}, and averaged geometric distances to the LMC~\cite{2022ApJ...934L...7R}, SMC~\cite{2024ApJ...973...30B}, M31~\cite{vilardell2010distance}, and NGC4258~\cite{2020ApJ...891L...1P}. 
At greater distances, the orange and red dots show extragalactic distance precision calibrated from the period-luminosity relationship of Cepheids and TRGB, respectively~\cite{2022ApJ...934L...7R,Freedman:2020dne,Freedman:2024eph}. The solid green line shows the typical distance precision of individual SN Ia whose absolute luminosity is calibrated by these same Cepheids~\cite{2022ApJ...934L...7R} or TRGB~\cite{Freedman:2024eph}.

\paragraph*{\bf CDL Calibration.---} EEM's precise SN distance measurements of Eq.~\eqref{eq:sigmaD} can be leveraged to improve the CDL and by extension $H_0$ inferences, either by providing a set of distant geometric anchors to calibrate Cepheids (or the TRGB), or by pinning down the standard absolute magnitude of SNe Ia directly.

The EEM can establish a set of geometric anchors more distant than the LMC, SMC, M31, or NGC4258~\cite{2020ApJ...891L...1P} galaxies---the first rung of the CDL. When used as anchors for Cepheids, each additional anchor
possesses uncertainties from the EEM $D_A$ measurement itself, as well as from the period-luminosity relationship ($\sigma_{\rm PL} \approx 0.4\%)$, reddening $(\sigma_R \approx 0.1\%)$, and metallicity $(\sigma_z \approx 0.2\%)$ of Cepheids within each anchor galaxy~\cite{Riess:2016jrr,2019ApJ...876...85R,2022ApJ...934L...7R}. The fractional error contribution to $H_0$ arising from a set of Type IIP SNe anchors with apparent magnitude $\{m_1, ..., m_n\}$ is then 
\begin{align}
\label{eq:anchorVariance}
    \left.{\sigma}_{\rm rung \,1,EEM}^2 = 1 \,
      \!\middle/\,
    \sum _{i=1}^n \frac{1}{\frac{\sigma_{D_A}}{D_A}(m_i,t_{{\rm obs,}i})^2 + \sigma_{\rm Cepheid}^2}  \right. \, ,
\end{align}
where $\sigma_{\rm Cepheid}^2 = \sigma_{\rm PL}^2 + \sigma_R^2 + \sigma_z^2 \approx (0.5\%)^2$ and $t_{{\rm obs,}i}$ is the observation time of the $i$th SN. The optimal observational strategy of allocating $t_{{\rm obs,}i}$, constrained to a fixed total $t_\mathrm{obs}$, to minimize Eq.~\eqref{eq:anchorVariance} can be found from a Lagrange multiplier method~(see App.~\ref{sec:optimumSearchStrategyCepheids}). 

The dark blue band in the right panel of Fig.~\ref{fig:systemsDistanceUncertaintyPlot} shows the minimized ${\sigma}_{\rm rung \, 1, EEM}$ for $t_\mathrm{obs} = 5 \, \mathrm{yr}$ of continuous operation (assuming a duty cycle of 25\%) as a function of Matchlight. More conservatively, the light blue band shows the minimum ${\sigma}_{\rm rung \, 1, EEM}$ if only the brightest SN IIP in 5 years of operation is observed for $t_{\rm obs} = 90 \times 6$ hours over its three-month duration. At low Matchlight, the lower boundary of the light- and dark-blue bands meet, demonstrating that in this regime the sensitivity is dominated by observing the single brightest SN in a typical 5-year observation window. 

The blue, orange, green, and purple horizontal lines respectively show the current SH0ES fractional uncertainties in $H_0$ from calibrating the CDL's rung 1, rung 2, rung 3, and known systematics~\cite{2022ApJ...934L...7R}. The last three contribute unavoidable errors to the fractional uncertainty in $H_0$ that must be added in quadrature to the EEM calibration of the Cepheids. Currently, this additional error contribution to $H_0$  is around $\sigma_{\rm rung \,2}^2 + \sigma_{\rm rung \, 3}^2 + \sigma_{\rm sys }^2 \simeq (1.15\%)^2$~\cite{2022ApJ...934L...7R}. The EEM calibration of Cepheids can resolve or confirm the Hubble tension at $n \sigma$ if $n \sqrt{\sigma_{\rm rung \,1,EEM}^2 + (1.15\%)^2} \lesssim 7\%$, which is the current discrepancy between the ACT+\textit{Planck}~\cite{2025arXiv250314452L} and SH0ES~\cite{2024ApJ...973...30B} values for $H_0$, as shown by the dashed red line. The right panel of Fig.~\ref{fig:systemsDistanceUncertaintyPlot} shows that interferometer arrays with $\mathrm{Matchlight} \gtrsim 400$ ($600$) possess sufficient statistical power to distinguish between the discrepant Hubble constant measurements at the $2\sigma$ ($3\sigma$) level, thereby offering a path to clarify the tension.

The EEM can also standardize the absolute magnitude of SNe Ia directly---the second rung of the CDL (\emph{skipping} the first). The error contribution in $H_0$ from this rung is
\begin{align}
\label{eq:SNIaVariance}
    \left.{\sigma}_{\rm rung,2}^2 = 1 \,
      \!\middle/\,
    \sum _{i=1}^n \frac{1}{\frac{\sigma_{D_A}}{D_A}(m_i,t_{{\rm obs,}i})^2 + \sigma_{\rm Ia}^2}  \right. \, ,
\end{align}
where $\sigma_{\rm Ia} \approx 0.13\, \rm mag \times \ln(10)/5 \simeq 6 \%$ is due to the intrinsic scatter in the SN Ia magnitudes~\cite{2022ApJ...934L...7R}. The orange band in Fig.~\ref{fig:systemsDistanceUncertaintyPlot} shows the minimum $\sigma_{\rm rung,2}$ for 5 years of continuous interferometer operation, similarly optimized via Lagrange multipliers (see App.~\ref{sec:optimumSearchStrategyCepheids}). The light orange band shows the minimum ${\sigma}_{\rm rung \, 2, EEM}$ if only the \emph{brightest} Type Ia SN in 5 years of operation is observed for $t_{\rm obs} = 30 \times 6$ hours (1 month of nightly observation). The band is nearly independent of Matchlight since the uncertainty is dominated by the intrinsic SN Ia magnitude scatter, $\sigma_{\rm Ia}$, for this singular SN. 

Analogous to the calibration of the first rung, the total error in $H_0$ from a direct EEM calibration of SNe Ia includes not only Eq.~\eqref{eq:SNIaVariance}, but also the quadrature sum of downstream uncertainties, $\sigma_{\rm rung,3}^2 + \sigma_{\rm sys}^2 \simeq (0.5\%)^2$. Under this criterion, the EEM approach can provide a meaningful test at $n\sigma$ of the Hubble tension if $n\sqrt{\sigma_{\rm rung,2,EEM}^2 + (0.5\%)^2} \lesssim 7\%$, with $n=2~(4)$ requiring $\mathrm{Matchlight} \gtrsim 100~ (400)$.

\paragraph*{\bf EEM-based $\boldsymbol{H_0}$ measurement.---}
The EEM can also determine $H_0$ directly by measuring distances to SNe within the Hubble flow, bypassing the traditional CDL entirely. The statistical uncertainty in $H_0$ from observing a set of SNe with apparent magnitudes $\{m_1,,,,m_n\}$ is
\begin{align}
\label{eq:H0statUncertainty}
    \left.\frac{{\sigma}_{H_0,\rm stat}^2}{H_0^2} = 1 \,
      \!\middle/\,
    \sum _{i=1}^n \frac{1}{\frac{\sigma_{D_A}}{D_A}(m_i,t_{{\rm obs,}i})^2 + \sigma_{v}^2 / v_i^2} \right. \, .
\end{align}
The term $\sigma_v^2 / v_i^2 \simeq \sigma_v^2 / H_0^2 D_L(m_i)^2$ reflects the contribution of peculiar velocity-induced redshift scatter, with host galaxies typically exhibiting velocity dispersions of \( \sigma_v \approx 250 \)~km/s relative to the Hubble flow~\cite{Pan-STARRS1:2017jku}.

\begin{figure}
    \centering
    \includegraphics[width=.497\textwidth]{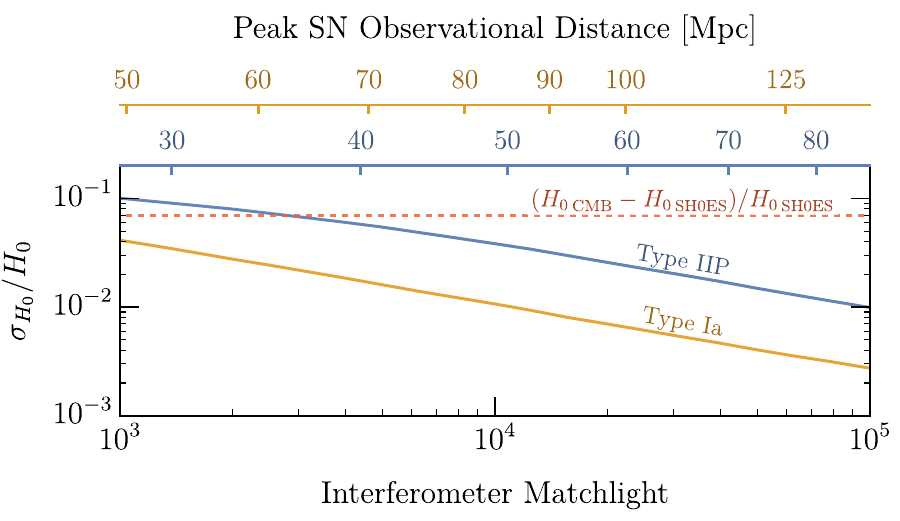}
    \caption{
    EEM-only fractional uncertainties on $H_0$ using SNe IIP (Ia) in blue (orange) as a function of an intensity interferometer's Matchlight (Eq.~\eqref{eq:matchlight}). The solid curves show the total uncertainty for a 5-yr campaign using the optimal observation strategy, including EEM-based statistical uncertainty, peculiar-velocity scatter, and cosmic variance. 
    The upper horizontal axes show the optimal SN distance. 
    The dashed red line depicts the discrepancy in $H_0$ between refs.~\cite{2025arXiv250314452L,2024ApJ...973...30B}.}
    \label{fig:directH0Plot}
\end{figure}

The optimal observation strategy to minimize Eq.~\eqref{eq:H0statUncertainty} balances observations of the more abundant distant, but dimmer, SNe that typically have larger fractional uncertainties in $D_A$ (but smaller in $v$), against the more rare, but brighter, SNe with very precise $D_A$ determinations but larger uncertainties due to peculiar velocity. 
Cosmic variance further contributes a systematic uncertainty ${\sigma_{H_0, \rm sys}/}{H_0} = \sigma_{\rm cv} (m)$, coming from measuring a local \emph{subvolume} of the Universe whose expansion rate may not be representative of the global value of $H_0$ due to local density fluctuations~\cite{Marra:2013rba,Odderskov:2017ivg} (cfr.~App.~\ref{sec:cosmicVariance}). 
To a good approximation, a near-optimal observational strategy entails observing SNe around an apparent magnitude optimum $m$ and corresponding $D_L(m)$ (upper axes in Fig.~\ref{fig:directH0Plot}) with a narrow spread with standard deviation $\sigma_m \approx 0.2$.
The resulting cosmic variance $\sigma_{\rm cv} (m)$ is then proportional to the mean density fluctuation within a sphere of radius $D_L(m)$ .
We prove this result in App.~\ref{sec:optimumSearchStrategyH0}, where we present a Lagrange multiplier method that allocates the optimal observation times for realistic SN populations, constrained to a given Matchlight and total observation time.

In Fig.~\ref{fig:directH0Plot}, we plot in blue (orange) the $H_0$ sensitivity of this optimal strategy for continuous intensity interferometer monitoring of Type IIP (Ia) SNe over a total observation time of 5~years. This direct EEM-based $H_0$ measurement is limited by the combination of peculiar-velocity scatter and cosmic variance, which both decrease with distance, and the rapid loss of $\mathrm{SNR} \propto 1/D_L^2$ (cfr.~Eq.~\eqref{eq:sigmaD}) at larger distances. Nevertheless, a highly capable intensity interferometer with $\mathrm{Matchlight} = 10^5$ could obtain a geometric $H_0$ inference at 1\% (0.3\%), independent of other methods.

\paragraph*{\bf Conclusions.---}  
We have investigated the EEM's potential as a novel, purely geometric technique for measuring SN distances by directly comparing their physical ejecta velocities to their angular expansion rates resolved via intensity interferometry (Eq.~\eqref{eq:D_A}, Fig.~\ref{fig:eemDiagram}). Unlike traditional luminosity-based approaches, the EEM does not rely on external flux calibration or assumptions about extinction, blackbody dilution, or symmetry. We have demonstrated three flagship applications: (1) geometric calibration of Cepheids using SNe IIP as moderate-distance anchors; (2) direct standardization of SNe Ia; and (3) construction of a fully independent Hubble diagram. The application of the EEM on \emph{multiple} lines simultaneously, and on SNe Ia (or other more exotic types) specifically, deserves further study, although ref.~\cite{2025arXiv250307725K} has already performed modeling along this direction. Moreover, SNe of different types have been observed in the \emph{same} galaxy in recent years~\cite{Soderberg:2008uh,SNesamegalaxy}, which can provide calibration cross-checks of Cepheids, TRGBs, and EEM modeling of SNe Ia. 

Under optimized campaigns with spectrally-multiplexed long-baseline intensity interferometers, our forecasts indicate that subpercent distance uncertainties are achievable for each application. Arrays with $\mathrm{Matchlight}\gtrsim 1{,}250$ can constrain the first-rung Cepheid calibration to better than $1\%$ over 5 years. Over the same observation time, direct SNe~Ia standardization (skipping the first rung entirely) should be possible to the $1\%$ level with $\mathrm{Matchlight}\gtrsim 1{,}250$. These CDL (re-)calibrations are sufficient to resolve or confirm at more than $4\sigma$ and $6\sigma$, respectively, any (unknown) systematics that may be responsible for the current $H_0$ tension. Moreover, by targeting SNe in the Hubble flow, EEM offers a ladder-free route to $H_0$ with competitive precision (Fig.~\ref{fig:directH0Plot}), providing a systematics-orthogonal cross-check to CMB, gravitational-wave standard sirens, strong-lensing time delays, and other emerging probes.  

Next-generation intensity interferometers---featuring longer baselines, larger collecting areas, improved timing resolution, and high-resolution spectral multiplexing---promise a multitude of new scientific applications~\cite{VanTilburg:2023tkl,Galanis:2023gef,Dalal:2024aaj}. In the long run, they could drive EEM uncertainties on $H_0$ below those of CMB measurements. Phase-retrieval and model-independent imaging techniques may soon enable direct reconstruction of SN ejecta morphology, further mitigating model biases. Coupled with wide-field transient surveys and coordinated spectral follow-ups, the EEM can anchor the cosmic distance scale on purely geometric grounds, opening a new era of precision cosmology.

\acknowledgments
We thank Masha Baryakhtar, Lars Bildsten, Neal Dalal, Marios Galanis, Jared Goldberg, and Yong-Zhong Qian for helpful discussions.
This material is based upon work supported by the National Science Foundation under Grant No.~PHY-2210551.
DD is supported by the James Arthur Postdoctoral Fellowship.
IC is supported by the James Arthur Graduate Associate Fellowship and the NYU GSAS Dissertation Writing Fellowship.
JH is grateful for the hospitality of NYU and CCA, where part of this work was carried out.

Research at Perimeter Institute is supported in part by the Government of Canada through the Department of Innovation, Science and Economic Development and by the Province of Ontario through the Ministry of Colleges and Universities. The Center for Computational Astrophysics at the Flatiron Institute is supported by the Simons Foundation. 

\bibliography{eem.bib}

\begin{thebibliography}{82}%
\makeatletter
\providecommand \@ifxundefined [1]{%
 \@ifx{#1\undefined}
}%
\providecommand \@ifnum [1]{%
 \ifnum #1\expandafter \@firstoftwo
 \else \expandafter \@secondoftwo
 \fi
}%
\providecommand \@ifx [1]{%
 \ifx #1\expandafter \@firstoftwo
 \else \expandafter \@secondoftwo
 \fi
}%
\providecommand \natexlab [1]{#1}%
\providecommand \enquote  [1]{``#1''}%
\providecommand \bibnamefont  [1]{#1}%
\providecommand \bibfnamefont [1]{#1}%
\providecommand \citenamefont [1]{#1}%
\providecommand \href@noop [0]{\@secondoftwo}%
\providecommand \href [0]{\begingroup \@sanitize@url \@href}%
\providecommand \@href[1]{\@@startlink{#1}\@@href}%
\providecommand \@@href[1]{\endgroup#1\@@endlink}%
\providecommand \@sanitize@url [0]{\catcode `\\12\catcode `\$12\catcode
  `\&12\catcode `\#12\catcode `\^12\catcode `\_12\catcode `\%12\relax}%
\providecommand \@@startlink[1]{}%
\providecommand \@@endlink[0]{}%
\providecommand \url  [0]{\begingroup\@sanitize@url \@url }%
\providecommand \@url [1]{\endgroup\@href {#1}{\urlprefix }}%
\providecommand \urlprefix  [0]{URL }%
\providecommand \Eprint [0]{\href }%
\providecommand \doibase [0]{https://doi.org/}%
\providecommand \selectlanguage [0]{\@gobble}%
\providecommand \bibinfo  [0]{\@secondoftwo}%
\providecommand \bibfield  [0]{\@secondoftwo}%
\providecommand \translation [1]{[#1]}%
\providecommand \BibitemOpen [0]{}%
\providecommand \bibitemStop [0]{}%
\providecommand \bibitemNoStop [0]{.\EOS\space}%
\providecommand \EOS [0]{\spacefactor3000\relax}%
\providecommand \BibitemShut  [1]{\csname bibitem#1\endcsname}%
\let\auto@bib@innerbib\@empty
\bibitem [{\citenamefont {Chen}\ \emph {et~al.}(2025)\citenamefont {Chen},
  \citenamefont {Dunsky}, \citenamefont {Tilburg}, \citenamefont {Huang},\ and\
  \citenamefont {Wagoner}}]{eem-1}%
  \BibitemOpen
  \bibfield  {author} {\bibinfo {author} {\bibfnamefont {I.-K.}\ \bibnamefont
  {Chen}}, \bibinfo {author} {\bibfnamefont {D.}~\bibnamefont {Dunsky}},
  \bibinfo {author} {\bibfnamefont {K.~V.}\ \bibnamefont {Tilburg}}, \bibinfo
  {author} {\bibfnamefont {J.}~\bibnamefont {Huang}},\ and\ \bibinfo {author}
  {\bibfnamefont {R.~V.}\ \bibnamefont {Wagoner}},\ }\href
  {https://arxiv.org/abs/2504.20132} {\bibinfo {title} {Expanding ejecta
  method: I. mapping supernova morphology with intensity interferometry}}
  (\bibinfo {year} {2025}),\ \Eprint {https://arxiv.org/abs/2504.20132}
  {arXiv:2504.20132 [astro-ph.HE]} \BibitemShut {NoStop}%
\bibitem [{\citenamefont {{Dark Energy Survey and Kilo-Degree Survey
  Collaboration}}\ \emph {et~al.}(2023)\citenamefont {{Dark Energy Survey and
  Kilo-Degree Survey Collaboration}} \emph {et~al.}}]{2023OJAp....6E..36D}%
  \BibitemOpen
  \bibfield  {author} {\bibinfo {author} {\bibnamefont {{Dark Energy Survey and
  Kilo-Degree Survey Collaboration}}} \emph {et~al.},\ }\bibfield  {title}
  {\bibinfo {title} {{DES Y3 + KiDS-1000: Consistent cosmology combining cosmic
  shear surveys}},\ }\href {https://doi.org/10.21105/astro.2305.17173}
  {\bibfield  {journal} {\bibinfo  {journal} {The Open Journal of
  Astrophysics}\ }\textbf {\bibinfo {volume} {6}},\ \bibinfo {eid} {36}
  (\bibinfo {year} {2023})},\ \Eprint {https://arxiv.org/abs/2305.17173}
  {arXiv:2305.17173 [astro-ph.CO]} \BibitemShut {NoStop}%
\bibitem [{\citenamefont {{Philcox}}\ and\ \citenamefont
  {{Ivanov}}(2022)}]{2022PhRvD.105d3517P}%
  \BibitemOpen
  \bibfield  {author} {\bibinfo {author} {\bibfnamefont {O.~H.~E.}\
  \bibnamefont {{Philcox}}}\ and\ \bibinfo {author} {\bibfnamefont {M.~M.}\
  \bibnamefont {{Ivanov}}},\ }\bibfield  {title} {\bibinfo {title} {{BOSS DR12
  full-shape cosmology: {\ensuremath{\Lambda}} CDM constraints from the
  large-scale galaxy power spectrum and bispectrum monopole}},\ }\href
  {https://doi.org/10.1103/PhysRevD.105.043517} {\bibfield  {journal} {\bibinfo
   {journal} {\prd}\ }\textbf {\bibinfo {volume} {105}},\ \bibinfo {eid}
  {043517} (\bibinfo {year} {2022})},\ \Eprint
  {https://arxiv.org/abs/2112.04515} {arXiv:2112.04515 [astro-ph.CO]}
  \BibitemShut {NoStop}%
\bibitem [{\citenamefont {{DESI Collaboration}}\ \emph
  {et~al.}(2024)\citenamefont {{DESI Collaboration}} \emph
  {et~al.}}]{2024arXiv241112022D}%
  \BibitemOpen
  \bibfield  {author} {\bibinfo {author} {\bibnamefont {{DESI Collaboration}}}
  \emph {et~al.},\ }\bibfield  {title} {\bibinfo {title} {{DESI 2024 VII:
  Cosmological Constraints from the Full-Shape Modeling of Clustering
  Measurements}},\ }\href {https://doi.org/10.48550/arXiv.2411.12022}
  {\bibfield  {journal} {\bibinfo  {journal} {arXiv e-prints}\ ,\ \bibinfo
  {eid} {arXiv:2411.12022}} (\bibinfo {year} {2024})},\ \Eprint
  {https://arxiv.org/abs/2411.12022} {arXiv:2411.12022 [astro-ph.CO]}
  \BibitemShut {NoStop}%
\bibitem [{\citenamefont {{DESI Collaboration}}\ \emph
  {et~al.}(2025)\citenamefont {{DESI Collaboration}} \emph
  {et~al.}}]{2025arXiv250314738D}%
  \BibitemOpen
  \bibfield  {author} {\bibinfo {author} {\bibnamefont {{DESI Collaboration}}}
  \emph {et~al.},\ }\bibfield  {title} {\bibinfo {title} {{DESI DR2 Results II:
  Measurements of Baryon Acoustic Oscillations and Cosmological Constraints}},\
  }\href@noop {} {\bibfield  {journal} {\bibinfo  {journal} {arXiv e-prints}\
  ,\ \bibinfo {eid} {arXiv:2503.14738}} (\bibinfo {year} {2025})},\ \Eprint
  {https://arxiv.org/abs/2503.14738} {arXiv:2503.14738 [astro-ph.CO]}
  \BibitemShut {NoStop}%
\bibitem [{\citenamefont {{Jungman}}\ \emph {et~al.}(1996)\citenamefont
  {{Jungman}}, \citenamefont {{Kamionkowski}}, \citenamefont {{Kosowsky}},\
  and\ \citenamefont {{Spergel}}}]{1996PhRvD..54.1332J}%
  \BibitemOpen
  \bibfield  {author} {\bibinfo {author} {\bibfnamefont {G.}~\bibnamefont
  {{Jungman}}}, \bibinfo {author} {\bibfnamefont {M.}~\bibnamefont
  {{Kamionkowski}}}, \bibinfo {author} {\bibfnamefont {A.}~\bibnamefont
  {{Kosowsky}}},\ and\ \bibinfo {author} {\bibfnamefont {D.~N.}\ \bibnamefont
  {{Spergel}}},\ }\bibfield  {title} {\bibinfo {title} {{Cosmological-parameter
  determination with microwave background maps}},\ }\href
  {https://doi.org/10.1103/PhysRevD.54.1332} {\bibfield  {journal} {\bibinfo
  {journal} {\prd}\ }\textbf {\bibinfo {volume} {54}},\ \bibinfo {pages} {1332}
  (\bibinfo {year} {1996})},\ \Eprint {https://arxiv.org/abs/astro-ph/9512139}
  {arXiv:astro-ph/9512139 [astro-ph]} \BibitemShut {NoStop}%
\bibitem [{\citenamefont {Aylor}\ \emph {et~al.}(2019)\citenamefont {Aylor},
  \citenamefont {Joy}, \citenamefont {Knox}, \citenamefont {Millea},
  \citenamefont {Raghunathan},\ and\ \citenamefont {Wu}}]{Aylor:2018drw}%
  \BibitemOpen
  \bibfield  {author} {\bibinfo {author} {\bibfnamefont {K.}~\bibnamefont
  {Aylor}}, \bibinfo {author} {\bibfnamefont {M.}~\bibnamefont {Joy}}, \bibinfo
  {author} {\bibfnamefont {L.}~\bibnamefont {Knox}}, \bibinfo {author}
  {\bibfnamefont {M.}~\bibnamefont {Millea}}, \bibinfo {author} {\bibfnamefont
  {S.}~\bibnamefont {Raghunathan}},\ and\ \bibinfo {author} {\bibfnamefont
  {W.~L.~K.}\ \bibnamefont {Wu}},\ }\bibfield  {title} {\bibinfo {title}
  {{Sounds Discordant: Classical Distance Ladder \& $\Lambda$CDM -based
  Determinations of the Cosmological Sound Horizon}},\ }\href
  {https://doi.org/10.3847/1538-4357/ab0898} {\bibfield  {journal} {\bibinfo
  {journal} {Astrophys. J.}\ }\textbf {\bibinfo {volume} {874}},\ \bibinfo
  {pages} {4} (\bibinfo {year} {2019})},\ \Eprint
  {https://arxiv.org/abs/1811.00537} {arXiv:1811.00537 [astro-ph.CO]}
  \BibitemShut {NoStop}%
\bibitem [{\citenamefont {{Madore}}\ and\ \citenamefont
  {{Freedman}}(1991)}]{1991PASP..103..933M}%
  \BibitemOpen
  \bibfield  {author} {\bibinfo {author} {\bibfnamefont {B.~F.}\ \bibnamefont
  {{Madore}}}\ and\ \bibinfo {author} {\bibfnamefont {W.~L.}\ \bibnamefont
  {{Freedman}}},\ }\bibfield  {title} {\bibinfo {title} {{The Cepheid Distance
  Scale}},\ }\href {https://doi.org/10.1086/132911} {\bibfield  {journal}
  {\bibinfo  {journal} {\pasp}\ }\textbf {\bibinfo {volume} {103}},\ \bibinfo
  {pages} {933} (\bibinfo {year} {1991})}\BibitemShut {NoStop}%
\bibitem [{\citenamefont {{Rizzi}}\ \emph {et~al.}(2007)\citenamefont
  {{Rizzi}}, \citenamefont {{Tully}}, \citenamefont {{Makarov}}, \citenamefont
  {{Makarova}}, \citenamefont {{Dolphin}}, \citenamefont {{Sakai}},\ and\
  \citenamefont {{Shaya}}}]{2007ApJ...661..815R}%
  \BibitemOpen
  \bibfield  {author} {\bibinfo {author} {\bibfnamefont {L.}~\bibnamefont
  {{Rizzi}}}, \bibinfo {author} {\bibfnamefont {R.~B.}\ \bibnamefont
  {{Tully}}}, \bibinfo {author} {\bibfnamefont {D.}~\bibnamefont {{Makarov}}},
  \bibinfo {author} {\bibfnamefont {L.}~\bibnamefont {{Makarova}}}, \bibinfo
  {author} {\bibfnamefont {A.~E.}\ \bibnamefont {{Dolphin}}}, \bibinfo {author}
  {\bibfnamefont {S.}~\bibnamefont {{Sakai}}},\ and\ \bibinfo {author}
  {\bibfnamefont {E.~J.}\ \bibnamefont {{Shaya}}},\ }\bibfield  {title}
  {\bibinfo {title} {{Tip of the Red Giant Branch Distances. II. Zero-Point
  Calibration}},\ }\href {https://doi.org/10.1086/516566} {\bibfield  {journal}
  {\bibinfo  {journal} {\apj}\ }\textbf {\bibinfo {volume} {661}},\ \bibinfo
  {pages} {815} (\bibinfo {year} {2007})},\ \Eprint
  {https://arxiv.org/abs/astro-ph/0701518} {arXiv:astro-ph/0701518 [astro-ph]}
  \BibitemShut {NoStop}%
\bibitem [{\citenamefont {{Riess}}\ \emph {et~al.}(2018)\citenamefont
  {{Riess}}, \citenamefont {{Casertano}}, \citenamefont {{Yuan}}, \citenamefont
  {{Macri}}, \citenamefont {{Bucciarelli}}, \citenamefont {{Lattanzi}},
  \citenamefont {{MacKenty}}, \citenamefont {{Bowers}}, \citenamefont
  {{Zheng}}, \citenamefont {{Filippenko}}, \citenamefont {{Huang}},\ and\
  \citenamefont {{Anderson}}}]{2018ApJ...861..126R}%
  \BibitemOpen
  \bibfield  {author} {\bibinfo {author} {\bibfnamefont {A.~G.}\ \bibnamefont
  {{Riess}}}, \bibinfo {author} {\bibfnamefont {S.}~\bibnamefont
  {{Casertano}}}, \bibinfo {author} {\bibfnamefont {W.}~\bibnamefont {{Yuan}}},
  \bibinfo {author} {\bibfnamefont {L.}~\bibnamefont {{Macri}}}, \bibinfo
  {author} {\bibfnamefont {B.}~\bibnamefont {{Bucciarelli}}}, \bibinfo {author}
  {\bibfnamefont {M.~G.}\ \bibnamefont {{Lattanzi}}}, \bibinfo {author}
  {\bibfnamefont {J.~W.}\ \bibnamefont {{MacKenty}}}, \bibinfo {author}
  {\bibfnamefont {J.~B.}\ \bibnamefont {{Bowers}}}, \bibinfo {author}
  {\bibfnamefont {W.}~\bibnamefont {{Zheng}}}, \bibinfo {author} {\bibfnamefont
  {A.~V.}\ \bibnamefont {{Filippenko}}}, \bibinfo {author} {\bibfnamefont
  {C.}~\bibnamefont {{Huang}}},\ and\ \bibinfo {author} {\bibfnamefont {R.~I.}\
  \bibnamefont {{Anderson}}},\ }\bibfield  {title} {\bibinfo {title} {{Milky
  Way Cepheid Standards for Measuring Cosmic Distances and Application to Gaia
  DR2: Implications for the Hubble Constant}},\ }\href
  {https://doi.org/10.3847/1538-4357/aac82e} {\bibfield  {journal} {\bibinfo
  {journal} {\apj}\ }\textbf {\bibinfo {volume} {861}},\ \bibinfo {eid} {126}
  (\bibinfo {year} {2018})},\ \Eprint {https://arxiv.org/abs/1804.10655}
  {arXiv:1804.10655 [astro-ph.CO]} \BibitemShut {NoStop}%
\bibitem [{\citenamefont {{Riess}}\ \emph {et~al.}(2021)\citenamefont
  {{Riess}}, \citenamefont {{Casertano}}, \citenamefont {{Yuan}}, \citenamefont
  {{Bowers}}, \citenamefont {{Macri}}, \citenamefont {{Zinn}},\ and\
  \citenamefont {{Scolnic}}}]{2021ApJ...908L...6R}%
  \BibitemOpen
  \bibfield  {author} {\bibinfo {author} {\bibfnamefont {A.~G.}\ \bibnamefont
  {{Riess}}}, \bibinfo {author} {\bibfnamefont {S.}~\bibnamefont
  {{Casertano}}}, \bibinfo {author} {\bibfnamefont {W.}~\bibnamefont {{Yuan}}},
  \bibinfo {author} {\bibfnamefont {J.~B.}\ \bibnamefont {{Bowers}}}, \bibinfo
  {author} {\bibfnamefont {L.}~\bibnamefont {{Macri}}}, \bibinfo {author}
  {\bibfnamefont {J.~C.}\ \bibnamefont {{Zinn}}},\ and\ \bibinfo {author}
  {\bibfnamefont {D.}~\bibnamefont {{Scolnic}}},\ }\bibfield  {title} {\bibinfo
  {title} {{Cosmic Distances Calibrated to 1\% Precision with Gaia EDR3
  Parallaxes and Hubble Space Telescope Photometry of 75 Milky Way Cepheids
  Confirm Tension with {\ensuremath{\Lambda}}CDM}},\ }\href
  {https://doi.org/10.3847/2041-8213/abdbaf} {\bibfield  {journal} {\bibinfo
  {journal} {\apjl}\ }\textbf {\bibinfo {volume} {908}},\ \bibinfo {eid} {L6}
  (\bibinfo {year} {2021})},\ \Eprint {https://arxiv.org/abs/2012.08534}
  {arXiv:2012.08534 [astro-ph.CO]} \BibitemShut {NoStop}%
\bibitem [{\citenamefont {{Riess}}\ \emph {et~al.}(2019)\citenamefont
  {{Riess}}, \citenamefont {{Casertano}}, \citenamefont {{Yuan}}, \citenamefont
  {{Macri}},\ and\ \citenamefont {{Scolnic}}}]{2019ApJ...876...85R}%
  \BibitemOpen
  \bibfield  {author} {\bibinfo {author} {\bibfnamefont {A.~G.}\ \bibnamefont
  {{Riess}}}, \bibinfo {author} {\bibfnamefont {S.}~\bibnamefont
  {{Casertano}}}, \bibinfo {author} {\bibfnamefont {W.}~\bibnamefont {{Yuan}}},
  \bibinfo {author} {\bibfnamefont {L.~M.}\ \bibnamefont {{Macri}}},\ and\
  \bibinfo {author} {\bibfnamefont {D.}~\bibnamefont {{Scolnic}}},\ }\bibfield
  {title} {\bibinfo {title} {{Large Magellanic Cloud Cepheid Standards Provide
  a 1\% Foundation for the Determination of the Hubble Constant and Stronger
  Evidence for Physics beyond {\ensuremath{\Lambda}}CDM}},\ }\href
  {https://doi.org/10.3847/1538-4357/ab1422} {\bibfield  {journal} {\bibinfo
  {journal} {\apj}\ }\textbf {\bibinfo {volume} {876}},\ \bibinfo {eid} {85}
  (\bibinfo {year} {2019})},\ \Eprint {https://arxiv.org/abs/1903.07603}
  {arXiv:1903.07603 [astro-ph.CO]} \BibitemShut {NoStop}%
\bibitem [{\citenamefont {{Pesce}}\ \emph {et~al.}(2020)\citenamefont
  {{Pesce}}, \citenamefont {{Braatz}}, \citenamefont {{Reid}}, \citenamefont
  {{Riess}}, \citenamefont {{Scolnic}}, \citenamefont {{Condon}}, \citenamefont
  {{Gao}}, \citenamefont {{Henkel}}, \citenamefont {{Impellizzeri}},
  \citenamefont {{Kuo}},\ and\ \citenamefont {{Lo}}}]{2020ApJ...891L...1P}%
  \BibitemOpen
  \bibfield  {author} {\bibinfo {author} {\bibfnamefont {D.~W.}\ \bibnamefont
  {{Pesce}}}, \bibinfo {author} {\bibfnamefont {J.~A.}\ \bibnamefont
  {{Braatz}}}, \bibinfo {author} {\bibfnamefont {M.~J.}\ \bibnamefont
  {{Reid}}}, \bibinfo {author} {\bibfnamefont {A.~G.}\ \bibnamefont {{Riess}}},
  \bibinfo {author} {\bibfnamefont {D.}~\bibnamefont {{Scolnic}}}, \bibinfo
  {author} {\bibfnamefont {J.~J.}\ \bibnamefont {{Condon}}}, \bibinfo {author}
  {\bibfnamefont {F.}~\bibnamefont {{Gao}}}, \bibinfo {author} {\bibfnamefont
  {C.}~\bibnamefont {{Henkel}}}, \bibinfo {author} {\bibfnamefont {C.~M.~V.}\
  \bibnamefont {{Impellizzeri}}}, \bibinfo {author} {\bibfnamefont {C.~Y.}\
  \bibnamefont {{Kuo}}},\ and\ \bibinfo {author} {\bibfnamefont {K.~Y.}\
  \bibnamefont {{Lo}}},\ }\bibfield  {title} {\bibinfo {title} {{The Megamaser
  Cosmology Project. XIII. Combined Hubble Constant Constraints}},\ }\href
  {https://doi.org/10.3847/2041-8213/ab75f0} {\bibfield  {journal} {\bibinfo
  {journal} {\apjl}\ }\textbf {\bibinfo {volume} {891}},\ \bibinfo {eid} {L1}
  (\bibinfo {year} {2020})},\ \Eprint {https://arxiv.org/abs/2001.09213}
  {arXiv:2001.09213 [astro-ph.CO]} \BibitemShut {NoStop}%
\bibitem [{\citenamefont {{Freedman}}\ and\ \citenamefont
  {{Madore}}(2010)}]{2010ARA&A..48..673F}%
  \BibitemOpen
  \bibfield  {author} {\bibinfo {author} {\bibfnamefont {W.~L.}\ \bibnamefont
  {{Freedman}}}\ and\ \bibinfo {author} {\bibfnamefont {B.~F.}\ \bibnamefont
  {{Madore}}},\ }\bibfield  {title} {\bibinfo {title} {{The Hubble Constant}},\
  }\href {https://doi.org/10.1146/annurev-astro-082708-101829} {\bibfield
  {journal} {\bibinfo  {journal} {\araa}\ }\textbf {\bibinfo {volume} {48}},\
  \bibinfo {pages} {673} (\bibinfo {year} {2010})},\ \Eprint
  {https://arxiv.org/abs/1004.1856} {arXiv:1004.1856 [astro-ph.CO]}
  \BibitemShut {NoStop}%
\bibitem [{\citenamefont {{Kamionkowski}}\ and\ \citenamefont
  {{Riess}}(2023)}]{2023ARNPS..73..153K}%
  \BibitemOpen
  \bibfield  {author} {\bibinfo {author} {\bibfnamefont {M.}~\bibnamefont
  {{Kamionkowski}}}\ and\ \bibinfo {author} {\bibfnamefont {A.~G.}\
  \bibnamefont {{Riess}}},\ }\bibfield  {title} {\bibinfo {title} {{The Hubble
  Tension and Early Dark Energy}},\ }\href
  {https://doi.org/10.1146/annurev-nucl-111422-024107} {\bibfield  {journal}
  {\bibinfo  {journal} {Annual Review of Nuclear and Particle Science}\
  }\textbf {\bibinfo {volume} {73}},\ \bibinfo {pages} {153} (\bibinfo {year}
  {2023})},\ \Eprint {https://arxiv.org/abs/2211.04492} {arXiv:2211.04492
  [astro-ph.CO]} \BibitemShut {NoStop}%
\bibitem [{\citenamefont {{Bennett}}\ \emph {et~al.}(2013)\citenamefont
  {{Bennett}} \emph {et~al.}}]{2013ApJS..208...20B}%
  \BibitemOpen
  \bibfield  {author} {\bibinfo {author} {\bibfnamefont {C.~L.}\ \bibnamefont
  {{Bennett}}} \emph {et~al.},\ }\bibfield  {title} {\bibinfo {title}
  {{Nine-year Wilkinson Microwave Anisotropy Probe (WMAP) Observations: Final
  Maps and Results}},\ }\href {https://doi.org/10.1088/0067-0049/208/2/20}
  {\bibfield  {journal} {\bibinfo  {journal} {\apjs}\ }\textbf {\bibinfo
  {volume} {208}},\ \bibinfo {eid} {20} (\bibinfo {year} {2013})},\ \Eprint
  {https://arxiv.org/abs/1212.5225} {arXiv:1212.5225 [astro-ph.CO]}
  \BibitemShut {NoStop}%
\bibitem [{\citenamefont {{Planck Collaboration}}\ \emph
  {et~al.}(2020)\citenamefont {{Planck Collaboration}} \emph
  {et~al.}}]{2020A&A...641A...6P}%
  \BibitemOpen
  \bibfield  {author} {\bibinfo {author} {\bibnamefont {{Planck
  Collaboration}}} \emph {et~al.},\ }\bibfield  {title} {\bibinfo {title}
  {{Planck 2018 results. VI. Cosmological parameters}},\ }\href
  {https://doi.org/10.1051/0004-6361/201833910} {\bibfield  {journal} {\bibinfo
   {journal} {\aap}\ }\textbf {\bibinfo {volume} {641}},\ \bibinfo {eid} {A6}
  (\bibinfo {year} {2020})},\ \Eprint {https://arxiv.org/abs/1807.06209}
  {arXiv:1807.06209 [astro-ph.CO]} \BibitemShut {NoStop}%
\bibitem [{\citenamefont {{SPT-3G Collaboration}}\ \emph
  {et~al.}(2021)\citenamefont {{SPT-3G Collaboration}} \emph
  {et~al.}}]{2021PhRvD.104b2003D}%
  \BibitemOpen
  \bibfield  {author} {\bibinfo {author} {\bibnamefont {{SPT-3G
  Collaboration}}} \emph {et~al.},\ }\bibfield  {title} {\bibinfo {title}
  {{Measurements of the E -mode polarization and temperature-E -mode
  correlation of the CMB from SPT-3G 2018 data}},\ }\href
  {https://doi.org/10.1103/PhysRevD.104.022003} {\bibfield  {journal} {\bibinfo
   {journal} {\prd}\ }\textbf {\bibinfo {volume} {104}},\ \bibinfo {eid}
  {022003} (\bibinfo {year} {2021})},\ \Eprint
  {https://arxiv.org/abs/2101.01684} {arXiv:2101.01684 [astro-ph.CO]}
  \BibitemShut {NoStop}%
\bibitem [{\citenamefont {{Aiola}}\ \emph {et~al.}(2020)\citenamefont {{Aiola}}
  \emph {et~al.}}]{2020JCAP...12..047A}%
  \BibitemOpen
  \bibfield  {author} {\bibinfo {author} {\bibfnamefont {S.}~\bibnamefont
  {{Aiola}}} \emph {et~al.},\ }\bibfield  {title} {\bibinfo {title} {{The
  Atacama Cosmology Telescope: DR4 maps and cosmological parameters}},\ }\href
  {https://doi.org/10.1088/1475-7516/2020/12/047} {\bibfield  {journal}
  {\bibinfo  {journal} {\jcap}\ }\textbf {\bibinfo {volume} {2020}},\ \bibinfo
  {eid} {047} (\bibinfo {year} {2020})},\ \Eprint
  {https://arxiv.org/abs/2007.07288} {arXiv:2007.07288 [astro-ph.CO]}
  \BibitemShut {NoStop}%
\bibitem [{\citenamefont {{Louis}}\ \emph {et~al.}(2025)\citenamefont {{Louis}}
  \emph {et~al.}}]{2025arXiv250314452L}%
  \BibitemOpen
  \bibfield  {author} {\bibinfo {author} {\bibfnamefont {T.}~\bibnamefont
  {{Louis}}} \emph {et~al.},\ }\bibfield  {title} {\bibinfo {title} {{The
  Atacama Cosmology Telescope: DR6 Power Spectra, Likelihoods and $\Lambda$CDM
  Parameters}},\ }\href {https://doi.org/10.48550/arXiv.2503.14452} {\bibfield
  {journal} {\bibinfo  {journal} {arXiv e-prints}\ ,\ \bibinfo {eid}
  {arXiv:2503.14452}} (\bibinfo {year} {2025})},\ \Eprint
  {https://arxiv.org/abs/2503.14452} {arXiv:2503.14452 [astro-ph.CO]}
  \BibitemShut {NoStop}%
\bibitem [{\citenamefont {{Freedman}}\ \emph {et~al.}(2024)\citenamefont
  {{Freedman}}, \citenamefont {{Madore}}, \citenamefont {{Jang}}, \citenamefont
  {{Hoyt}}, \citenamefont {{Lee}},\ and\ \citenamefont
  {{Owens}}}]{2024arXiv240806153F}%
  \BibitemOpen
  \bibfield  {author} {\bibinfo {author} {\bibfnamefont {W.~L.}\ \bibnamefont
  {{Freedman}}}, \bibinfo {author} {\bibfnamefont {B.~F.}\ \bibnamefont
  {{Madore}}}, \bibinfo {author} {\bibfnamefont {I.~S.}\ \bibnamefont
  {{Jang}}}, \bibinfo {author} {\bibfnamefont {T.~J.}\ \bibnamefont {{Hoyt}}},
  \bibinfo {author} {\bibfnamefont {A.~J.}\ \bibnamefont {{Lee}}},\ and\
  \bibinfo {author} {\bibfnamefont {K.~A.}\ \bibnamefont {{Owens}}},\
  }\bibfield  {title} {\bibinfo {title} {{Status Report on the Chicago-Carnegie
  Hubble Program (CCHP): Measurement of the Hubble Constant Using the Hubble
  and James Webb Space Telescopes}},\ }\href
  {https://doi.org/10.48550/arXiv.2408.06153} {\bibfield  {journal} {\bibinfo
  {journal} {arXiv e-prints}\ ,\ \bibinfo {eid} {arXiv:2408.06153}} (\bibinfo
  {year} {2024})},\ \Eprint {https://arxiv.org/abs/2408.06153}
  {arXiv:2408.06153 [astro-ph.CO]} \BibitemShut {NoStop}%
\bibitem [{\citenamefont {{Riess}}\ \emph {et~al.}(2022)\citenamefont
  {{Riess}}, \citenamefont {{Yuan}}, \citenamefont {{Macri}}, \citenamefont
  {{Scolnic}}, \citenamefont {{Brout}}, \citenamefont {{Casertano}},
  \citenamefont {{Jones}}, \citenamefont {{Murakami}}, \citenamefont {{Anand}},
  \citenamefont {{Breuval}}, \citenamefont {{Brink}}, \citenamefont
  {{Filippenko}}, \citenamefont {{Hoffmann}}, \citenamefont {{Jha}},
  \citenamefont {{D'arcy Kenworthy}}, \citenamefont {{Mackenty}}, \citenamefont
  {{Stahl}},\ and\ \citenamefont {{Zheng}}}]{2022ApJ...934L...7R}%
  \BibitemOpen
  \bibfield  {author} {\bibinfo {author} {\bibfnamefont {A.~G.}\ \bibnamefont
  {{Riess}}}, \bibinfo {author} {\bibfnamefont {W.}~\bibnamefont {{Yuan}}},
  \bibinfo {author} {\bibfnamefont {L.~M.}\ \bibnamefont {{Macri}}}, \bibinfo
  {author} {\bibfnamefont {D.}~\bibnamefont {{Scolnic}}}, \bibinfo {author}
  {\bibfnamefont {D.}~\bibnamefont {{Brout}}}, \bibinfo {author} {\bibfnamefont
  {S.}~\bibnamefont {{Casertano}}}, \bibinfo {author} {\bibfnamefont {D.~O.}\
  \bibnamefont {{Jones}}}, \bibinfo {author} {\bibfnamefont {Y.}~\bibnamefont
  {{Murakami}}}, \bibinfo {author} {\bibfnamefont {G.~S.}\ \bibnamefont
  {{Anand}}}, \bibinfo {author} {\bibfnamefont {L.}~\bibnamefont {{Breuval}}},
  \bibinfo {author} {\bibfnamefont {T.~G.}\ \bibnamefont {{Brink}}}, \bibinfo
  {author} {\bibfnamefont {A.~V.}\ \bibnamefont {{Filippenko}}}, \bibinfo
  {author} {\bibfnamefont {S.}~\bibnamefont {{Hoffmann}}}, \bibinfo {author}
  {\bibfnamefont {S.~W.}\ \bibnamefont {{Jha}}}, \bibinfo {author}
  {\bibfnamefont {W.}~\bibnamefont {{D'arcy Kenworthy}}}, \bibinfo {author}
  {\bibfnamefont {J.}~\bibnamefont {{Mackenty}}}, \bibinfo {author}
  {\bibfnamefont {B.~E.}\ \bibnamefont {{Stahl}}},\ and\ \bibinfo {author}
  {\bibfnamefont {W.}~\bibnamefont {{Zheng}}},\ }\bibfield  {title} {\bibinfo
  {title} {{A Comprehensive Measurement of the Local Value of the Hubble
  Constant with 1 km s$^{-1}$ Mpc$^{-1}$ Uncertainty from the Hubble Space
  Telescope and the SH0ES Team}},\ }\href
  {https://doi.org/10.3847/2041-8213/ac5c5b} {\bibfield  {journal} {\bibinfo
  {journal} {\apjl}\ }\textbf {\bibinfo {volume} {934}},\ \bibinfo {eid} {L7}
  (\bibinfo {year} {2022})},\ \Eprint {https://arxiv.org/abs/2112.04510}
  {arXiv:2112.04510 [astro-ph.CO]} \BibitemShut {NoStop}%
\bibitem [{\citenamefont {{Breuval}}\ \emph {et~al.}(2024)\citenamefont
  {{Breuval}}, \citenamefont {{Riess}}, \citenamefont {{Casertano}},
  \citenamefont {{Yuan}}, \citenamefont {{Macri}}, \citenamefont
  {{Romaniello}}, \citenamefont {{Murakami}}, \citenamefont {{Scolnic}},
  \citenamefont {{Anand}},\ and\ \citenamefont
  {{Soszy{\'n}ski}}}]{2024ApJ...973...30B}%
  \BibitemOpen
  \bibfield  {author} {\bibinfo {author} {\bibfnamefont {L.}~\bibnamefont
  {{Breuval}}}, \bibinfo {author} {\bibfnamefont {A.~G.}\ \bibnamefont
  {{Riess}}}, \bibinfo {author} {\bibfnamefont {S.}~\bibnamefont
  {{Casertano}}}, \bibinfo {author} {\bibfnamefont {W.}~\bibnamefont {{Yuan}}},
  \bibinfo {author} {\bibfnamefont {L.~M.}\ \bibnamefont {{Macri}}}, \bibinfo
  {author} {\bibfnamefont {M.}~\bibnamefont {{Romaniello}}}, \bibinfo {author}
  {\bibfnamefont {Y.~S.}\ \bibnamefont {{Murakami}}}, \bibinfo {author}
  {\bibfnamefont {D.}~\bibnamefont {{Scolnic}}}, \bibinfo {author}
  {\bibfnamefont {G.~S.}\ \bibnamefont {{Anand}}},\ and\ \bibinfo {author}
  {\bibfnamefont {I.}~\bibnamefont {{Soszy{\'n}ski}}},\ }\bibfield  {title}
  {\bibinfo {title} {{Small Magellanic Cloud Cepheids Observed with the Hubble
  Space Telescope Provide a New Anchor for the SH0ES Distance Ladder}},\ }\href
  {https://doi.org/10.3847/1538-4357/ad630e} {\bibfield  {journal} {\bibinfo
  {journal} {\apj}\ }\textbf {\bibinfo {volume} {973}},\ \bibinfo {eid} {30}
  (\bibinfo {year} {2024})},\ \Eprint {https://arxiv.org/abs/2404.08038}
  {arXiv:2404.08038 [astro-ph.CO]} \BibitemShut {NoStop}%
\bibitem [{\citenamefont {{Schutz}}(1986)}]{1986Natur.323..310S}%
  \BibitemOpen
  \bibfield  {author} {\bibinfo {author} {\bibfnamefont {B.~F.}\ \bibnamefont
  {{Schutz}}},\ }\bibfield  {title} {\bibinfo {title} {{Determining the Hubble
  constant from gravitational wave observations}},\ }\href
  {https://doi.org/10.1038/323310a0} {\bibfield  {journal} {\bibinfo  {journal}
  {\nat}\ }\textbf {\bibinfo {volume} {323}},\ \bibinfo {pages} {310} (\bibinfo
  {year} {1986})}\BibitemShut {NoStop}%
\bibitem [{\citenamefont {Oguri}(2016)}]{Oguri:2016dgk}%
  \BibitemOpen
  \bibfield  {author} {\bibinfo {author} {\bibfnamefont {M.}~\bibnamefont
  {Oguri}},\ }\bibfield  {title} {\bibinfo {title} {{Measuring the
  distance-redshift relation with the cross-correlation of gravitational wave
  standard sirens and galaxies}},\ }\href
  {https://doi.org/10.1103/PhysRevD.93.083511} {\bibfield  {journal} {\bibinfo
  {journal} {Phys. Rev. D}\ }\textbf {\bibinfo {volume} {93}},\ \bibinfo
  {pages} {083511} (\bibinfo {year} {2016})},\ \Eprint
  {https://arxiv.org/abs/1603.02356} {arXiv:1603.02356 [astro-ph.CO]}
  \BibitemShut {NoStop}%
\bibitem [{\citenamefont {{Chen}}\ \emph {et~al.}(2018)\citenamefont {{Chen}},
  \citenamefont {{Fishbach}},\ and\ \citenamefont
  {{Holz}}}]{2018Natur.562..545C}%
  \BibitemOpen
  \bibfield  {author} {\bibinfo {author} {\bibfnamefont {H.-Y.}\ \bibnamefont
  {{Chen}}}, \bibinfo {author} {\bibfnamefont {M.}~\bibnamefont {{Fishbach}}},\
  and\ \bibinfo {author} {\bibfnamefont {D.~E.}\ \bibnamefont {{Holz}}},\
  }\bibfield  {title} {\bibinfo {title} {{A two per cent Hubble constant
  measurement from standard sirens within five years}},\ }\href
  {https://doi.org/10.1038/s41586-018-0606-0} {\bibfield  {journal} {\bibinfo
  {journal} {\nat}\ }\textbf {\bibinfo {volume} {562}},\ \bibinfo {pages} {545}
  (\bibinfo {year} {2018})},\ \Eprint {https://arxiv.org/abs/1712.06531}
  {arXiv:1712.06531 [astro-ph.CO]} \BibitemShut {NoStop}%
\bibitem [{\citenamefont {{Tonry}}\ and\ \citenamefont
  {{Schneider}}(1988)}]{1988AJ.....96..807T}%
  \BibitemOpen
  \bibfield  {author} {\bibinfo {author} {\bibfnamefont {J.}~\bibnamefont
  {{Tonry}}}\ and\ \bibinfo {author} {\bibfnamefont {D.~P.}\ \bibnamefont
  {{Schneider}}},\ }\bibfield  {title} {\bibinfo {title} {{A New Technique for
  Measuring Extragalactic Distances}},\ }\href {https://doi.org/10.1086/114847}
  {\bibfield  {journal} {\bibinfo  {journal} {\aj}\ }\textbf {\bibinfo {volume}
  {96}},\ \bibinfo {pages} {807} (\bibinfo {year} {1988})}\BibitemShut
  {NoStop}%
\bibitem [{\citenamefont {{Huang}}\ \emph {et~al.}(2020)\citenamefont
  {{Huang}}, \citenamefont {{Riess}}, \citenamefont {{Yuan}}, \citenamefont
  {{Macri}}, \citenamefont {{Zakamska}}, \citenamefont {{Casertano}},
  \citenamefont {{Whitelock}}, \citenamefont {{Hoffmann}}, \citenamefont
  {{Filippenko}},\ and\ \citenamefont {{Scolnic}}}]{2020ApJ...889....5H}%
  \BibitemOpen
  \bibfield  {author} {\bibinfo {author} {\bibfnamefont {C.~D.}\ \bibnamefont
  {{Huang}}}, \bibinfo {author} {\bibfnamefont {A.~G.}\ \bibnamefont
  {{Riess}}}, \bibinfo {author} {\bibfnamefont {W.}~\bibnamefont {{Yuan}}},
  \bibinfo {author} {\bibfnamefont {L.~M.}\ \bibnamefont {{Macri}}}, \bibinfo
  {author} {\bibfnamefont {N.~L.}\ \bibnamefont {{Zakamska}}}, \bibinfo
  {author} {\bibfnamefont {S.}~\bibnamefont {{Casertano}}}, \bibinfo {author}
  {\bibfnamefont {P.~A.}\ \bibnamefont {{Whitelock}}}, \bibinfo {author}
  {\bibfnamefont {S.~L.}\ \bibnamefont {{Hoffmann}}}, \bibinfo {author}
  {\bibfnamefont {A.~V.}\ \bibnamefont {{Filippenko}}},\ and\ \bibinfo {author}
  {\bibfnamefont {D.}~\bibnamefont {{Scolnic}}},\ }\bibfield  {title} {\bibinfo
  {title} {{Hubble Space Telescope Observations of Mira Variables in the SN Ia
  Host NGC 1559: An Alternative Candle to Measure the Hubble Constant}},\
  }\href {https://doi.org/10.3847/1538-4357/ab5dbd} {\bibfield  {journal}
  {\bibinfo  {journal} {\apj}\ }\textbf {\bibinfo {volume} {889}},\ \bibinfo
  {eid} {5} (\bibinfo {year} {2020})},\ \Eprint
  {https://arxiv.org/abs/1908.10883} {arXiv:1908.10883 [astro-ph.CO]}
  \BibitemShut {NoStop}%
\bibitem [{\citenamefont {{Refsdal}}(1964)}]{1964MNRAS.128..307R}%
  \BibitemOpen
  \bibfield  {author} {\bibinfo {author} {\bibfnamefont {S.}~\bibnamefont
  {{Refsdal}}},\ }\bibfield  {title} {\bibinfo {title} {{On the possibility of
  determining Hubble's parameter and the masses of galaxies from the
  gravitational lens effect}},\ }\href
  {https://doi.org/10.1093/mnras/128.4.307} {\bibfield  {journal} {\bibinfo
  {journal} {\mnras}\ }\textbf {\bibinfo {volume} {128}},\ \bibinfo {pages}
  {307} (\bibinfo {year} {1964})}\BibitemShut {NoStop}%
\bibitem [{\citenamefont {{Birrer}}\ \emph {et~al.}(2020)\citenamefont
  {{Birrer}}, \citenamefont {{Shajib}}, \citenamefont {{Galan}}, \citenamefont
  {{Millon}}, \citenamefont {{Treu}}, \citenamefont {{Agnello}}, \citenamefont
  {{Auger}}, \citenamefont {{Chen}}, \citenamefont {{Christensen}},
  \citenamefont {{Collett}}, \citenamefont {{Courbin}}, \citenamefont
  {{Fassnacht}}, \citenamefont {{Koopmans}}, \citenamefont {{Marshall}},
  \citenamefont {{Park}}, \citenamefont {{Rusu}}, \citenamefont {{Sluse}},
  \citenamefont {{Spiniello}}, \citenamefont {{Suyu}}, \citenamefont
  {{Wagner-Carena}}, \citenamefont {{Wong}}, \citenamefont {{Barnab{\`e}}},
  \citenamefont {{Bolton}}, \citenamefont {{Czoske}}, \citenamefont {{Ding}},
  \citenamefont {{Frieman}},\ and\ \citenamefont {{Van de
  Vyvere}}}]{2020A&A...643A.165B}%
  \BibitemOpen
  \bibfield  {author} {\bibinfo {author} {\bibfnamefont {S.}~\bibnamefont
  {{Birrer}}}, \bibinfo {author} {\bibfnamefont {A.~J.}\ \bibnamefont
  {{Shajib}}}, \bibinfo {author} {\bibfnamefont {A.}~\bibnamefont {{Galan}}},
  \bibinfo {author} {\bibfnamefont {M.}~\bibnamefont {{Millon}}}, \bibinfo
  {author} {\bibfnamefont {T.}~\bibnamefont {{Treu}}}, \bibinfo {author}
  {\bibfnamefont {A.}~\bibnamefont {{Agnello}}}, \bibinfo {author}
  {\bibfnamefont {M.}~\bibnamefont {{Auger}}}, \bibinfo {author} {\bibfnamefont
  {G.~C.~F.}\ \bibnamefont {{Chen}}}, \bibinfo {author} {\bibfnamefont
  {L.}~\bibnamefont {{Christensen}}}, \bibinfo {author} {\bibfnamefont
  {T.}~\bibnamefont {{Collett}}}, \bibinfo {author} {\bibfnamefont
  {F.}~\bibnamefont {{Courbin}}}, \bibinfo {author} {\bibfnamefont {C.~D.}\
  \bibnamefont {{Fassnacht}}}, \bibinfo {author} {\bibfnamefont {L.~V.~E.}\
  \bibnamefont {{Koopmans}}}, \bibinfo {author} {\bibfnamefont {P.~J.}\
  \bibnamefont {{Marshall}}}, \bibinfo {author} {\bibfnamefont {J.~W.}\
  \bibnamefont {{Park}}}, \bibinfo {author} {\bibfnamefont {C.~E.}\
  \bibnamefont {{Rusu}}}, \bibinfo {author} {\bibfnamefont {D.}~\bibnamefont
  {{Sluse}}}, \bibinfo {author} {\bibfnamefont {C.}~\bibnamefont
  {{Spiniello}}}, \bibinfo {author} {\bibfnamefont {S.~H.}\ \bibnamefont
  {{Suyu}}}, \bibinfo {author} {\bibfnamefont {S.}~\bibnamefont
  {{Wagner-Carena}}}, \bibinfo {author} {\bibfnamefont {K.~C.}\ \bibnamefont
  {{Wong}}}, \bibinfo {author} {\bibfnamefont {M.}~\bibnamefont
  {{Barnab{\`e}}}}, \bibinfo {author} {\bibfnamefont {A.~S.}\ \bibnamefont
  {{Bolton}}}, \bibinfo {author} {\bibfnamefont {O.}~\bibnamefont {{Czoske}}},
  \bibinfo {author} {\bibfnamefont {X.}~\bibnamefont {{Ding}}}, \bibinfo
  {author} {\bibfnamefont {J.~A.}\ \bibnamefont {{Frieman}}},\ and\ \bibinfo
  {author} {\bibfnamefont {L.}~\bibnamefont {{Van de Vyvere}}},\ }\bibfield
  {title} {\bibinfo {title} {{TDCOSMO. IV. Hierarchical time-delay cosmography
  - joint inference of the Hubble constant and galaxy density profiles}},\
  }\href {https://doi.org/10.1051/0004-6361/202038861} {\bibfield  {journal}
  {\bibinfo  {journal} {\aap}\ }\textbf {\bibinfo {volume} {643}},\ \bibinfo
  {eid} {A165} (\bibinfo {year} {2020})},\ \Eprint
  {https://arxiv.org/abs/2007.02941} {arXiv:2007.02941 [astro-ph.CO]}
  \BibitemShut {NoStop}%
\bibitem [{\citenamefont {{Boone}}\ and\ \citenamefont
  {{McQuinn}}(2023)}]{2023ApJ...947L..23B}%
  \BibitemOpen
  \bibfield  {author} {\bibinfo {author} {\bibfnamefont {K.}~\bibnamefont
  {{Boone}}}\ and\ \bibinfo {author} {\bibfnamefont {M.}~\bibnamefont
  {{McQuinn}}},\ }\bibfield  {title} {\bibinfo {title} {{Solar System-scale
  Interferometry on Fast Radio Bursts Could Measure Cosmic Distances with
  Subpercent Precision}},\ }\href {https://doi.org/10.3847/2041-8213/acc947}
  {\bibfield  {journal} {\bibinfo  {journal} {\apjl}\ }\textbf {\bibinfo
  {volume} {947}},\ \bibinfo {eid} {L23} (\bibinfo {year} {2023})},\ \Eprint
  {https://arxiv.org/abs/2210.07159} {arXiv:2210.07159 [astro-ph.CO]}
  \BibitemShut {NoStop}%
\bibitem [{\citenamefont {Dalal}\ \emph {et~al.}(2024)\citenamefont {Dalal},
  \citenamefont {Galanis}, \citenamefont {Gammie}, \citenamefont {Gralla},\
  and\ \citenamefont {Murray}}]{Dalal:2024aaj}%
  \BibitemOpen
  \bibfield  {author} {\bibinfo {author} {\bibfnamefont {N.}~\bibnamefont
  {Dalal}}, \bibinfo {author} {\bibfnamefont {M.}~\bibnamefont {Galanis}},
  \bibinfo {author} {\bibfnamefont {C.}~\bibnamefont {Gammie}}, \bibinfo
  {author} {\bibfnamefont {S.~E.}\ \bibnamefont {Gralla}},\ and\ \bibinfo
  {author} {\bibfnamefont {N.}~\bibnamefont {Murray}},\ }\bibfield  {title}
  {\bibinfo {title} {{Probing H0 and resolving AGN disks with ultrafast photon
  counters}},\ }\href {https://doi.org/10.1103/PhysRevD.109.123029} {\bibfield
  {journal} {\bibinfo  {journal} {Phys. Rev. D}\ }\textbf {\bibinfo {volume}
  {109}},\ \bibinfo {pages} {123029} (\bibinfo {year} {2024})},\ \Eprint
  {https://arxiv.org/abs/2403.15903} {arXiv:2403.15903 [astro-ph.CO]}
  \BibitemShut {NoStop}%
\bibitem [{\citenamefont {{Kirshner}}\ and\ \citenamefont
  {{Kwan}}(1974)}]{KirshnerKwan}%
  \BibitemOpen
  \bibfield  {author} {\bibinfo {author} {\bibfnamefont {R.~P.}\ \bibnamefont
  {{Kirshner}}}\ and\ \bibinfo {author} {\bibfnamefont {J.}~\bibnamefont
  {{Kwan}}},\ }\bibfield  {title} {\bibinfo {title} {{Distances to
  extragalactic supernovae.}},\ }\href {https://doi.org/10.1086/153123}
  {\bibfield  {journal} {\bibinfo  {journal} {\apj}\ }\textbf {\bibinfo
  {volume} {193}},\ \bibinfo {pages} {27} (\bibinfo {year} {1974})}\BibitemShut
  {NoStop}%
\bibitem [{\citenamefont {{Schmidt}}\ \emph {et~al.}(1992)\citenamefont
  {{Schmidt}}, \citenamefont {{Kirshner}},\ and\ \citenamefont
  {{Eastman}}}]{1992ApJ...395..366S}%
  \BibitemOpen
  \bibfield  {author} {\bibinfo {author} {\bibfnamefont {B.~P.}\ \bibnamefont
  {{Schmidt}}}, \bibinfo {author} {\bibfnamefont {R.~P.}\ \bibnamefont
  {{Kirshner}}},\ and\ \bibinfo {author} {\bibfnamefont {R.~G.}\ \bibnamefont
  {{Eastman}}},\ }\bibfield  {title} {\bibinfo {title} {{Expanding Photospheres
  of Type II Supernovae and the Extragalactic Distance Scale}},\ }\href
  {https://doi.org/10.1086/171659} {\bibfield  {journal} {\bibinfo  {journal}
  {\apj}\ }\textbf {\bibinfo {volume} {395}},\ \bibinfo {pages} {366} (\bibinfo
  {year} {1992})},\ \Eprint {https://arxiv.org/abs/astro-ph/9204004}
  {arXiv:astro-ph/9204004 [astro-ph]} \BibitemShut {NoStop}%
\bibitem [{\citenamefont {{Kirshner}}\ \emph {et~al.}(1973)\citenamefont
  {{Kirshner}}, \citenamefont {{Oke}}, \citenamefont {{Penston}},\ and\
  \citenamefont {{Searle}}}]{1973ApJ...185..303K}%
  \BibitemOpen
  \bibfield  {author} {\bibinfo {author} {\bibfnamefont {R.~P.}\ \bibnamefont
  {{Kirshner}}}, \bibinfo {author} {\bibfnamefont {J.~B.}\ \bibnamefont
  {{Oke}}}, \bibinfo {author} {\bibfnamefont {M.~V.}\ \bibnamefont
  {{Penston}}},\ and\ \bibinfo {author} {\bibfnamefont {L.}~\bibnamefont
  {{Searle}}},\ }\bibfield  {title} {\bibinfo {title} {{The spectra of
  supernovae.}},\ }\href {https://doi.org/10.1086/152417} {\bibfield  {journal}
  {\bibinfo  {journal} {\apj}\ }\textbf {\bibinfo {volume} {185}},\ \bibinfo
  {pages} {303} (\bibinfo {year} {1973})}\BibitemShut {NoStop}%
\bibitem [{\citenamefont {{Schmidt}}\ \emph {et~al.}(1994)\citenamefont
  {{Schmidt}}, \citenamefont {{Kirshner}}, \citenamefont {{Eastman}},
  \citenamefont {{Phillips}}, \citenamefont {{Suntzeff}}, \citenamefont
  {{Hamuy}}, \citenamefont {{Maza}},\ and\ \citenamefont
  {{Aviles}}}]{1994ApJ...432...42S}%
  \BibitemOpen
  \bibfield  {author} {\bibinfo {author} {\bibfnamefont {B.~P.}\ \bibnamefont
  {{Schmidt}}}, \bibinfo {author} {\bibfnamefont {R.~P.}\ \bibnamefont
  {{Kirshner}}}, \bibinfo {author} {\bibfnamefont {R.~G.}\ \bibnamefont
  {{Eastman}}}, \bibinfo {author} {\bibfnamefont {M.~M.}\ \bibnamefont
  {{Phillips}}}, \bibinfo {author} {\bibfnamefont {N.~B.}\ \bibnamefont
  {{Suntzeff}}}, \bibinfo {author} {\bibfnamefont {M.}~\bibnamefont {{Hamuy}}},
  \bibinfo {author} {\bibfnamefont {J.}~\bibnamefont {{Maza}}},\ and\ \bibinfo
  {author} {\bibfnamefont {R.}~\bibnamefont {{Aviles}}},\ }\bibfield  {title}
  {\bibinfo {title} {{The Distance of Five Type II Supernovae Using the
  Expanding Photosphere Method and the Value of H 0}},\ }\href
  {https://doi.org/10.1086/174546} {\bibfield  {journal} {\bibinfo  {journal}
  {\apj}\ }\textbf {\bibinfo {volume} {432}},\ \bibinfo {pages} {42} (\bibinfo
  {year} {1994})}\BibitemShut {NoStop}%
\bibitem [{\citenamefont {{Wagoner}}(1981)}]{1981ApJ...250L..65W}%
  \BibitemOpen
  \bibfield  {author} {\bibinfo {author} {\bibfnamefont {R.~V.}\ \bibnamefont
  {{Wagoner}}},\ }\bibfield  {title} {\bibinfo {title} {{Effects of scattering
  on continuous radiation from supernovae and determination of their
  distances}},\ }\href {https://doi.org/10.1086/183675} {\bibfield  {journal}
  {\bibinfo  {journal} {\apjl}\ }\textbf {\bibinfo {volume} {250}},\ \bibinfo
  {pages} {L65} (\bibinfo {year} {1981})}\BibitemShut {NoStop}%
\bibitem [{\citenamefont {{Wagoner}}\ and\ \citenamefont
  {{Montes}}(1993)}]{1993PhR...227..205W}%
  \BibitemOpen
  \bibfield  {author} {\bibinfo {author} {\bibfnamefont {R.~V.}\ \bibnamefont
  {{Wagoner}}}\ and\ \bibinfo {author} {\bibfnamefont {M.}~\bibnamefont
  {{Montes}}},\ }\bibfield  {title} {\bibinfo {title} {{Cosmological distances
  from supernova photospheres}},\ }\href
  {https://doi.org/10.1016/0370-1573(93)90066-M} {\bibfield  {journal}
  {\bibinfo  {journal} {\physrep}\ }\textbf {\bibinfo {volume} {227}},\
  \bibinfo {pages} {205} (\bibinfo {year} {1993})}\BibitemShut {NoStop}%
\bibitem [{\citenamefont {{Sim}}(2017)}]{2017hsn..book..769S}%
  \BibitemOpen
  \bibfield  {author} {\bibinfo {author} {\bibfnamefont {S.~A.}\ \bibnamefont
  {{Sim}}},\ }\bibfield  {title} {\bibinfo {title} {{Spectra of Supernovae
  During the Photospheric Phase}},\ }in\ \href
  {https://doi.org/10.1007/978-3-319-21846-5_28} {\emph {\bibinfo {booktitle}
  {Handbook of Supernovae}}},\ \bibinfo {editor} {edited by\ \bibinfo {editor}
  {\bibfnamefont {A.~W.}\ \bibnamefont {{Alsabti}}}\ and\ \bibinfo {editor}
  {\bibfnamefont {P.}~\bibnamefont {{Murdin}}}}\ (\bibinfo {year} {2017})\ p.\
  \bibinfo {pages} {769}\BibitemShut {NoStop}%
\bibitem [{\citenamefont {Filippenko}(1997)}]{filippenko1997optical}%
  \BibitemOpen
  \bibfield  {author} {\bibinfo {author} {\bibfnamefont {A.~V.}\ \bibnamefont
  {Filippenko}},\ }\bibfield  {title} {\bibinfo {title} {Optical spectra of
  supernovae},\ }\href@noop {} {\bibfield  {journal} {\bibinfo  {journal}
  {Annual Review of Astronomy and Astrophysics}\ }\textbf {\bibinfo {volume}
  {35}},\ \bibinfo {pages} {309} (\bibinfo {year} {1997})}\BibitemShut
  {NoStop}%
\bibitem [{\citenamefont {{Mitchell}}\ \emph {et~al.}(2023)\citenamefont
  {{Mitchell}}, \citenamefont {{Didier}}, \citenamefont {{Ganesh}},
  \citenamefont {{Acharya}}, \citenamefont {{Khadka}},\ and\ \citenamefont
  {{Silwal}}}]{2023ApJ...942...38M}%
  \BibitemOpen
  \bibfield  {author} {\bibinfo {author} {\bibfnamefont {R.~C.}\ \bibnamefont
  {{Mitchell}}}, \bibinfo {author} {\bibfnamefont {B.}~\bibnamefont
  {{Didier}}}, \bibinfo {author} {\bibfnamefont {S.}~\bibnamefont {{Ganesh}}},
  \bibinfo {author} {\bibfnamefont {K.}~\bibnamefont {{Acharya}}}, \bibinfo
  {author} {\bibfnamefont {R.}~\bibnamefont {{Khadka}}},\ and\ \bibinfo
  {author} {\bibfnamefont {B.}~\bibnamefont {{Silwal}}},\ }\bibfield  {title}
  {\bibinfo {title} {{Locating Type II-P Supernovae Using the Expanding
  Photosphere Method. I. Comparing Distances from Different Line Velocities}},\
  }\href {https://doi.org/10.3847/1538-4357/aca415} {\bibfield  {journal}
  {\bibinfo  {journal} {\apj}\ }\textbf {\bibinfo {volume} {942}},\ \bibinfo
  {eid} {38} (\bibinfo {year} {2023})}\BibitemShut {NoStop}%
\bibitem [{\citenamefont {Bartel}\ \emph {et~al.}(2007)\citenamefont {Bartel},
  \citenamefont {Bietenholz}, \citenamefont {Rupen},\ and\ \citenamefont
  {Dwarkadas}}]{Bartel:2007px}%
  \BibitemOpen
  \bibfield  {author} {\bibinfo {author} {\bibfnamefont {N.}~\bibnamefont
  {Bartel}}, \bibinfo {author} {\bibfnamefont {M.~F.}\ \bibnamefont
  {Bietenholz}}, \bibinfo {author} {\bibfnamefont {M.~P.}\ \bibnamefont
  {Rupen}},\ and\ \bibinfo {author} {\bibfnamefont {V.~V.}\ \bibnamefont
  {Dwarkadas}},\ }\bibfield  {title} {\bibinfo {title} {{SN 1993J VLBI. 4. A
  Geometric Determination of the Distance to M81 with the Expanding Shock Front
  Method}},\ }\href {https://doi.org/10.1086/521326} {\bibfield  {journal}
  {\bibinfo  {journal} {Astrophys. J.}\ }\textbf {\bibinfo {volume} {668}},\
  \bibinfo {pages} {924} (\bibinfo {year} {2007})},\ \Eprint
  {https://arxiv.org/abs/0707.0881} {arXiv:0707.0881 [astro-ph]} \BibitemShut
  {NoStop}%
\bibitem [{\citenamefont {{Baron}}\ \emph {et~al.}(1995)\citenamefont
  {{Baron}}, \citenamefont {{Hauschildt}}, \citenamefont {{Branch}},
  \citenamefont {{Austin}}, \citenamefont {{Garnavich}}, \citenamefont {{Ann}},
  \citenamefont {{Wagner}}, \citenamefont {{Filippenko}}, \citenamefont
  {{Matheson}},\ and\ \citenamefont {{Liebert}}}]{1995ApJ...441..170B}%
  \BibitemOpen
  \bibfield  {author} {\bibinfo {author} {\bibfnamefont {E.}~\bibnamefont
  {{Baron}}}, \bibinfo {author} {\bibfnamefont {P.~H.}\ \bibnamefont
  {{Hauschildt}}}, \bibinfo {author} {\bibfnamefont {D.}~\bibnamefont
  {{Branch}}}, \bibinfo {author} {\bibfnamefont {S.}~\bibnamefont {{Austin}}},
  \bibinfo {author} {\bibfnamefont {P.}~\bibnamefont {{Garnavich}}}, \bibinfo
  {author} {\bibfnamefont {H.~B.}\ \bibnamefont {{Ann}}}, \bibinfo {author}
  {\bibfnamefont {R.~M.}\ \bibnamefont {{Wagner}}}, \bibinfo {author}
  {\bibfnamefont {A.~V.}\ \bibnamefont {{Filippenko}}}, \bibinfo {author}
  {\bibfnamefont {T.}~\bibnamefont {{Matheson}}},\ and\ \bibinfo {author}
  {\bibfnamefont {J.}~\bibnamefont {{Liebert}}},\ }\bibfield  {title} {\bibinfo
  {title} {{Non-LTE Spectral Analysis and Model Constraints on SN 1993J}},\
  }\href {https://doi.org/10.1086/175347} {\bibfield  {journal} {\bibinfo
  {journal} {\apj}\ }\textbf {\bibinfo {volume} {441}},\ \bibinfo {pages} {170}
  (\bibinfo {year} {1995})}\BibitemShut {NoStop}%
\bibitem [{\citenamefont {{Baron}}\ \emph {et~al.}(2004)\citenamefont
  {{Baron}}, \citenamefont {{Nugent}}, \citenamefont {{Branch}},\ and\
  \citenamefont {{Hauschildt}}}]{2004ApJ...616L..91B}%
  \BibitemOpen
  \bibfield  {author} {\bibinfo {author} {\bibfnamefont {E.}~\bibnamefont
  {{Baron}}}, \bibinfo {author} {\bibfnamefont {P.~E.}\ \bibnamefont
  {{Nugent}}}, \bibinfo {author} {\bibfnamefont {D.}~\bibnamefont {{Branch}}},\
  and\ \bibinfo {author} {\bibfnamefont {P.~H.}\ \bibnamefont {{Hauschildt}}},\
  }\bibfield  {title} {\bibinfo {title} {{Type IIP Supernovae as Cosmological
  Probes: A Spectral-fitting Expanding Atmosphere Model Distance to SN
  1999em}},\ }\href {https://doi.org/10.1086/426506} {\bibfield  {journal}
  {\bibinfo  {journal} {\apjl}\ }\textbf {\bibinfo {volume} {616}},\ \bibinfo
  {pages} {L91} (\bibinfo {year} {2004})},\ \Eprint
  {https://arxiv.org/abs/astro-ph/0410153} {arXiv:astro-ph/0410153 [astro-ph]}
  \BibitemShut {NoStop}%
\bibitem [{\citenamefont {{Hamuy}}\ and\ \citenamefont
  {{Pinto}}(2002)}]{2002ApJ...566L..63H}%
  \BibitemOpen
  \bibfield  {author} {\bibinfo {author} {\bibfnamefont {M.}~\bibnamefont
  {{Hamuy}}}\ and\ \bibinfo {author} {\bibfnamefont {P.~A.}\ \bibnamefont
  {{Pinto}}},\ }\bibfield  {title} {\bibinfo {title} {{Type II Supernovae as
  Standardized Candles}},\ }\href {https://doi.org/10.1086/339676} {\bibfield
  {journal} {\bibinfo  {journal} {\apjl}\ }\textbf {\bibinfo {volume} {566}},\
  \bibinfo {pages} {L63} (\bibinfo {year} {2002})},\ \Eprint
  {https://arxiv.org/abs/astro-ph/0201279} {arXiv:astro-ph/0201279 [astro-ph]}
  \BibitemShut {NoStop}%
\bibitem [{\citenamefont {{Rodr{\'\i}guez}}\ \emph {et~al.}(2014)\citenamefont
  {{Rodr{\'\i}guez}}, \citenamefont {{Clocchiatti}},\ and\ \citenamefont
  {{Hamuy}}}]{2014AJ....148..107R}%
  \BibitemOpen
  \bibfield  {author} {\bibinfo {author} {\bibfnamefont {{\'O}.}~\bibnamefont
  {{Rodr{\'\i}guez}}}, \bibinfo {author} {\bibfnamefont {A.}~\bibnamefont
  {{Clocchiatti}}},\ and\ \bibinfo {author} {\bibfnamefont {M.}~\bibnamefont
  {{Hamuy}}},\ }\bibfield  {title} {\bibinfo {title} {{Photospheric Magnitude
  Diagrams for Type II Supernovae: A Promising Tool to Compute Distances}},\
  }\href {https://doi.org/10.1088/0004-6256/148/6/107} {\bibfield  {journal}
  {\bibinfo  {journal} {\aj}\ }\textbf {\bibinfo {volume} {148}},\ \bibinfo
  {eid} {107} (\bibinfo {year} {2014})},\ \Eprint
  {https://arxiv.org/abs/1409.3198} {arXiv:1409.3198 [astro-ph.CO]}
  \BibitemShut {NoStop}%
\bibitem [{\citenamefont {{de Jaeger}}\ \emph {et~al.}(2015)\citenamefont {{de
  Jaeger}}, \citenamefont {{Gonz{\'a}lez-Gait{\'a}n}}, \citenamefont
  {{Anderson}}, \citenamefont {{Galbany}}, \citenamefont {{Hamuy}},
  \citenamefont {{Phillips}}, \citenamefont {{Stritzinger}}, \citenamefont
  {{Guti{\'e}rrez}}, \citenamefont {{Bolt}}, \citenamefont {{Burns}},
  \citenamefont {{Campillay}}, \citenamefont {{Castell{\'o}n}}, \citenamefont
  {{Contreras}}, \citenamefont {{Folatelli}}, \citenamefont {{Freedman}},
  \citenamefont {{Hsiao}}, \citenamefont {{Krisciunas}}, \citenamefont
  {{Krzeminski}}, \citenamefont {{Kuncarayakti}}, \citenamefont {{Morrell}},
  \citenamefont {{Olivares E.}}, \citenamefont {{Persson}},\ and\ \citenamefont
  {{Suntzeff}}}]{2015ApJ...815..121D}%
  \BibitemOpen
  \bibfield  {author} {\bibinfo {author} {\bibfnamefont {T.}~\bibnamefont {{de
  Jaeger}}}, \bibinfo {author} {\bibfnamefont {S.}~\bibnamefont
  {{Gonz{\'a}lez-Gait{\'a}n}}}, \bibinfo {author} {\bibfnamefont {J.~P.}\
  \bibnamefont {{Anderson}}}, \bibinfo {author} {\bibfnamefont
  {L.}~\bibnamefont {{Galbany}}}, \bibinfo {author} {\bibfnamefont
  {M.}~\bibnamefont {{Hamuy}}}, \bibinfo {author} {\bibfnamefont {M.~M.}\
  \bibnamefont {{Phillips}}}, \bibinfo {author} {\bibfnamefont {M.~D.}\
  \bibnamefont {{Stritzinger}}}, \bibinfo {author} {\bibfnamefont {C.~P.}\
  \bibnamefont {{Guti{\'e}rrez}}}, \bibinfo {author} {\bibfnamefont
  {L.}~\bibnamefont {{Bolt}}}, \bibinfo {author} {\bibfnamefont {C.~R.}\
  \bibnamefont {{Burns}}}, \bibinfo {author} {\bibfnamefont {A.}~\bibnamefont
  {{Campillay}}}, \bibinfo {author} {\bibfnamefont {S.}~\bibnamefont
  {{Castell{\'o}n}}}, \bibinfo {author} {\bibfnamefont {C.}~\bibnamefont
  {{Contreras}}}, \bibinfo {author} {\bibfnamefont {G.}~\bibnamefont
  {{Folatelli}}}, \bibinfo {author} {\bibfnamefont {W.~L.}\ \bibnamefont
  {{Freedman}}}, \bibinfo {author} {\bibfnamefont {E.~Y.}\ \bibnamefont
  {{Hsiao}}}, \bibinfo {author} {\bibfnamefont {K.}~\bibnamefont
  {{Krisciunas}}}, \bibinfo {author} {\bibfnamefont {W.}~\bibnamefont
  {{Krzeminski}}}, \bibinfo {author} {\bibfnamefont {H.}~\bibnamefont
  {{Kuncarayakti}}}, \bibinfo {author} {\bibfnamefont {N.}~\bibnamefont
  {{Morrell}}}, \bibinfo {author} {\bibfnamefont {F.}~\bibnamefont {{Olivares
  E.}}}, \bibinfo {author} {\bibfnamefont {S.~E.}\ \bibnamefont {{Persson}}},\
  and\ \bibinfo {author} {\bibfnamefont {N.}~\bibnamefont {{Suntzeff}}},\
  }\bibfield  {title} {\bibinfo {title} {{A Hubble Diagram from Type II
  Supernovae Based Solely on Photometry: The Photometric Color Method}},\
  }\href {https://doi.org/10.1088/0004-637X/815/2/121} {\bibfield  {journal}
  {\bibinfo  {journal} {\apj}\ }\textbf {\bibinfo {volume} {815}},\ \bibinfo
  {eid} {121} (\bibinfo {year} {2015})},\ \Eprint
  {https://arxiv.org/abs/1511.05145} {arXiv:1511.05145 [astro-ph.HE]}
  \BibitemShut {NoStop}%
\bibitem [{\citenamefont {{Sneppen}}\ \emph {et~al.}(2023)\citenamefont
  {{Sneppen}}, \citenamefont {{Watson}}, \citenamefont {{Poznanski}},
  \citenamefont {{Just}}, \citenamefont {{Bauswein}},\ and\ \citenamefont
  {{Wojtak}}}]{2023A&A...678A..14S}%
  \BibitemOpen
  \bibfield  {author} {\bibinfo {author} {\bibfnamefont {A.}~\bibnamefont
  {{Sneppen}}}, \bibinfo {author} {\bibfnamefont {D.}~\bibnamefont {{Watson}}},
  \bibinfo {author} {\bibfnamefont {D.}~\bibnamefont {{Poznanski}}}, \bibinfo
  {author} {\bibfnamefont {O.}~\bibnamefont {{Just}}}, \bibinfo {author}
  {\bibfnamefont {A.}~\bibnamefont {{Bauswein}}},\ and\ \bibinfo {author}
  {\bibfnamefont {R.}~\bibnamefont {{Wojtak}}},\ }\bibfield  {title} {\bibinfo
  {title} {{Measuring the Hubble constant with kilonovae using the expanding
  photosphere method}},\ }\href {https://doi.org/10.1051/0004-6361/202346306}
  {\bibfield  {journal} {\bibinfo  {journal} {\aap}\ }\textbf {\bibinfo
  {volume} {678}},\ \bibinfo {eid} {A14} (\bibinfo {year} {2023})},\ \Eprint
  {https://arxiv.org/abs/2306.12468} {arXiv:2306.12468 [astro-ph.CO]}
  \BibitemShut {NoStop}%
\bibitem [{\citenamefont {{Vogl}}\ \emph {et~al.}(2024)\citenamefont {{Vogl}},
  \citenamefont {{Taubenberger}}, \citenamefont {{Cs{\"o}rnyei}}, \citenamefont
  {{Leibundgut}}, \citenamefont {{Kerzendorf}}, \citenamefont {{Sim}},
  \citenamefont {{Blondin}}, \citenamefont {{Fl{\"o}rs}}, \citenamefont
  {{Holas}}, \citenamefont {{Shields}}, \citenamefont {{Spyromilio}},
  \citenamefont {{Suyu}},\ and\ \citenamefont
  {{Hillebrandt}}}]{2024arXiv241104968V}%
  \BibitemOpen
  \bibfield  {author} {\bibinfo {author} {\bibfnamefont {C.}~\bibnamefont
  {{Vogl}}}, \bibinfo {author} {\bibfnamefont {S.}~\bibnamefont
  {{Taubenberger}}}, \bibinfo {author} {\bibfnamefont {G.}~\bibnamefont
  {{Cs{\"o}rnyei}}}, \bibinfo {author} {\bibfnamefont {B.}~\bibnamefont
  {{Leibundgut}}}, \bibinfo {author} {\bibfnamefont {W.~E.}\ \bibnamefont
  {{Kerzendorf}}}, \bibinfo {author} {\bibfnamefont {S.~A.}\ \bibnamefont
  {{Sim}}}, \bibinfo {author} {\bibfnamefont {S.}~\bibnamefont {{Blondin}}},
  \bibinfo {author} {\bibfnamefont {A.}~\bibnamefont {{Fl{\"o}rs}}}, \bibinfo
  {author} {\bibfnamefont {A.}~\bibnamefont {{Holas}}}, \bibinfo {author}
  {\bibfnamefont {J.~V.}\ \bibnamefont {{Shields}}}, \bibinfo {author}
  {\bibfnamefont {J.}~\bibnamefont {{Spyromilio}}}, \bibinfo {author}
  {\bibfnamefont {S.~H.}\ \bibnamefont {{Suyu}}},\ and\ \bibinfo {author}
  {\bibfnamefont {W.}~\bibnamefont {{Hillebrandt}}},\ }\bibfield  {title}
  {\bibinfo {title} {{No rungs attached: A distance-ladder free determination
  of the Hubble constant through type II supernova spectral modelling}},\
  }\href {https://doi.org/10.48550/arXiv.2411.04968} {\bibfield  {journal}
  {\bibinfo  {journal} {arXiv e-prints}\ ,\ \bibinfo {eid} {arXiv:2411.04968}}
  (\bibinfo {year} {2024})},\ \Eprint {https://arxiv.org/abs/2411.04968}
  {arXiv:2411.04968 [astro-ph.CO]} \BibitemShut {NoStop}%
\bibitem [{\citenamefont {{Eastman}}\ \emph {et~al.}(1996)\citenamefont
  {{Eastman}}, \citenamefont {{Schmidt}},\ and\ \citenamefont
  {{Kirshner}}}]{1996ApJ...466..911E}%
  \BibitemOpen
  \bibfield  {author} {\bibinfo {author} {\bibfnamefont {R.~G.}\ \bibnamefont
  {{Eastman}}}, \bibinfo {author} {\bibfnamefont {B.~P.}\ \bibnamefont
  {{Schmidt}}},\ and\ \bibinfo {author} {\bibfnamefont {R.}~\bibnamefont
  {{Kirshner}}},\ }\bibfield  {title} {\bibinfo {title} {{The Atmospheres of
  Type II Supernovae and the Expanding Photosphere Method}},\ }\href
  {https://doi.org/10.1086/177563} {\bibfield  {journal} {\bibinfo  {journal}
  {\apj}\ }\textbf {\bibinfo {volume} {466}},\ \bibinfo {pages} {911} (\bibinfo
  {year} {1996})}\BibitemShut {NoStop}%
\bibitem [{\citenamefont {{Dessart}}\ and\ \citenamefont
  {{Hillier}}(2005)}]{2005A&A...439..671D}%
  \BibitemOpen
  \bibfield  {author} {\bibinfo {author} {\bibfnamefont {L.}~\bibnamefont
  {{Dessart}}}\ and\ \bibinfo {author} {\bibfnamefont {D.~J.}\ \bibnamefont
  {{Hillier}}},\ }\bibfield  {title} {\bibinfo {title} {{Distance
  determinations using type II supernovae and the expanding photosphere
  method}},\ }\href {https://doi.org/10.1051/0004-6361:20053217} {\bibfield
  {journal} {\bibinfo  {journal} {\aap}\ }\textbf {\bibinfo {volume} {439}},\
  \bibinfo {pages} {671} (\bibinfo {year} {2005})},\ \Eprint
  {https://arxiv.org/abs/astro-ph/0505465} {arXiv:astro-ph/0505465 [astro-ph]}
  \BibitemShut {NoStop}%
\bibitem [{\citenamefont {{Dessart}}\ and\ \citenamefont
  {{Hillier}}(2011)}]{2011MNRAS.415.3497D}%
  \BibitemOpen
  \bibfield  {author} {\bibinfo {author} {\bibfnamefont {L.}~\bibnamefont
  {{Dessart}}}\ and\ \bibinfo {author} {\bibfnamefont {D.~J.}\ \bibnamefont
  {{Hillier}}},\ }\bibfield  {title} {\bibinfo {title} {{Synthetic line and
  continuum linear-polarization signatures of axisymmetric Type II supernova
  ejecta}},\ }\href {https://doi.org/10.1111/j.1365-2966.2011.18967.x}
  {\bibfield  {journal} {\bibinfo  {journal} {\mnras}\ }\textbf {\bibinfo
  {volume} {415}},\ \bibinfo {pages} {3497} (\bibinfo {year} {2011})},\ \Eprint
  {https://arxiv.org/abs/1104.5346} {arXiv:1104.5346 [astro-ph.SR]}
  \BibitemShut {NoStop}%
\bibitem [{\citenamefont {{Dessart}}\ \emph {et~al.}(2015)\citenamefont
  {{Dessart}}, \citenamefont {{Hillier}}, \citenamefont {{Woosley}},
  \citenamefont {{Livne}}, \citenamefont {{Waldman}}, \citenamefont {{Yoon}},\
  and\ \citenamefont {{Langer}}}]{2015MNRAS.453.2189D}%
  \BibitemOpen
  \bibfield  {author} {\bibinfo {author} {\bibfnamefont {L.}~\bibnamefont
  {{Dessart}}}, \bibinfo {author} {\bibfnamefont {D.~J.}\ \bibnamefont
  {{Hillier}}}, \bibinfo {author} {\bibfnamefont {S.}~\bibnamefont
  {{Woosley}}}, \bibinfo {author} {\bibfnamefont {E.}~\bibnamefont {{Livne}}},
  \bibinfo {author} {\bibfnamefont {R.}~\bibnamefont {{Waldman}}}, \bibinfo
  {author} {\bibfnamefont {S.-C.}\ \bibnamefont {{Yoon}}},\ and\ \bibinfo
  {author} {\bibfnamefont {N.}~\bibnamefont {{Langer}}},\ }\bibfield  {title}
  {\bibinfo {title} {{Radiative-transfer models for supernovae IIb/Ib/Ic from
  binary-star progenitors}},\ }\href {https://doi.org/10.1093/mnras/stv1747}
  {\bibfield  {journal} {\bibinfo  {journal} {\mnras}\ }\textbf {\bibinfo
  {volume} {453}},\ \bibinfo {pages} {2189} (\bibinfo {year} {2015})},\ \Eprint
  {https://arxiv.org/abs/1507.07783} {arXiv:1507.07783 [astro-ph.SR]}
  \BibitemShut {NoStop}%
\bibitem [{\citenamefont {{Dessart}}\ \emph {et~al.}(2024)\citenamefont
  {{Dessart}}, \citenamefont {{Hillier}},\ and\ \citenamefont
  {{Leonard}}}]{2024A&A...684A..16D}%
  \BibitemOpen
  \bibfield  {author} {\bibinfo {author} {\bibfnamefont {L.}~\bibnamefont
  {{Dessart}}}, \bibinfo {author} {\bibfnamefont {D.~J.}\ \bibnamefont
  {{Hillier}}},\ and\ \bibinfo {author} {\bibfnamefont {D.~C.}\ \bibnamefont
  {{Leonard}}},\ }\bibfield  {title} {\bibinfo {title} {{The evolution of
  continuum polarization in type II supernovae as a diagnostic of ejecta
  morphology}},\ }\href {https://doi.org/10.1051/0004-6361/202347808}
  {\bibfield  {journal} {\bibinfo  {journal} {\aap}\ }\textbf {\bibinfo
  {volume} {684}},\ \bibinfo {eid} {A16} (\bibinfo {year} {2024})},\ \Eprint
  {https://arxiv.org/abs/2401.07330} {arXiv:2401.07330 [astro-ph.SR]}
  \BibitemShut {NoStop}%
\bibitem [{\citenamefont {{Dravins}}\ \emph {et~al.}(2015)\citenamefont
  {{Dravins}}, \citenamefont {{Lagadec}},\ and\ \citenamefont
  {{Nu{\~n}ez}}}]{2015A&A...580A..99D}%
  \BibitemOpen
  \bibfield  {author} {\bibinfo {author} {\bibfnamefont {D.}~\bibnamefont
  {{Dravins}}}, \bibinfo {author} {\bibfnamefont {T.}~\bibnamefont
  {{Lagadec}}},\ and\ \bibinfo {author} {\bibfnamefont {P.~D.}\ \bibnamefont
  {{Nu{\~n}ez}}},\ }\bibfield  {title} {\bibinfo {title} {{Long-baseline
  optical intensity interferometry. Laboratory demonstration of
  diffraction-limited imaging}},\ }\href
  {https://doi.org/10.1051/0004-6361/201526334} {\bibfield  {journal} {\bibinfo
   {journal} {\aap}\ }\textbf {\bibinfo {volume} {580}},\ \bibinfo {eid} {A99}
  (\bibinfo {year} {2015})},\ \Eprint {https://arxiv.org/abs/1506.05804}
  {arXiv:1506.05804 [astro-ph.IM]} \BibitemShut {NoStop}%
\bibitem [{\citenamefont {Dai}\ and\ \citenamefont {Dalal}(2025)}]{IIAI}%
  \BibitemOpen
  \bibfield  {author} {\bibinfo {author} {\bibfnamefont {B.}~\bibnamefont
  {Dai}}\ and\ \bibinfo {author} {\bibfnamefont {N.}~\bibnamefont {Dalal}}}
  (\bibinfo {year} {2025}),\ \bibinfo {note} {in preparation}\BibitemShut
  {NoStop}%
\bibitem [{\citenamefont {{Kim}}\ \emph {et~al.}(2025)\citenamefont {{Kim}},
  \citenamefont {{Nugent}}, \citenamefont {{Chen}}, \citenamefont {{Wang}},\
  and\ \citenamefont {{O'Brien}}}]{2025arXiv250307725K}%
  \BibitemOpen
  \bibfield  {author} {\bibinfo {author} {\bibfnamefont {A.~G.}\ \bibnamefont
  {{Kim}}}, \bibinfo {author} {\bibfnamefont {P.~E.}\ \bibnamefont {{Nugent}}},
  \bibinfo {author} {\bibfnamefont {X.}~\bibnamefont {{Chen}}}, \bibinfo
  {author} {\bibfnamefont {L.}~\bibnamefont {{Wang}}},\ and\ \bibinfo {author}
  {\bibfnamefont {J.~T.}\ \bibnamefont {{O'Brien}}},\ }\bibfield  {title}
  {\bibinfo {title} {{Measuring Type Ia Supernova Angular-Diameter Distances
  with Intensity Interferometry}},\ }\href
  {https://doi.org/10.48550/arXiv.2503.07725} {\bibfield  {journal} {\bibinfo
  {journal} {arXiv e-prints}\ ,\ \bibinfo {eid} {arXiv:2503.07725}} (\bibinfo
  {year} {2025})},\ \Eprint {https://arxiv.org/abs/2503.07725}
  {arXiv:2503.07725 [astro-ph.IM]} \BibitemShut {NoStop}%
\bibitem [{\citenamefont {Li}\ \emph {et~al.}(2011)\citenamefont {Li},
  \citenamefont {Leaman}, \citenamefont {Chornock}, \citenamefont {Filippenko},
  \citenamefont {Poznanski}, \citenamefont {Ganeshalingam}, \citenamefont
  {Wang}, \citenamefont {Modjaz}, \citenamefont {Jha}, \citenamefont {Foley}
  \emph {et~al.}}]{li2011nearby}%
  \BibitemOpen
  \bibfield  {author} {\bibinfo {author} {\bibfnamefont {W.}~\bibnamefont
  {Li}}, \bibinfo {author} {\bibfnamefont {J.}~\bibnamefont {Leaman}}, \bibinfo
  {author} {\bibfnamefont {R.}~\bibnamefont {Chornock}}, \bibinfo {author}
  {\bibfnamefont {A.~V.}\ \bibnamefont {Filippenko}}, \bibinfo {author}
  {\bibfnamefont {D.}~\bibnamefont {Poznanski}}, \bibinfo {author}
  {\bibfnamefont {M.}~\bibnamefont {Ganeshalingam}}, \bibinfo {author}
  {\bibfnamefont {X.}~\bibnamefont {Wang}}, \bibinfo {author} {\bibfnamefont
  {M.}~\bibnamefont {Modjaz}}, \bibinfo {author} {\bibfnamefont
  {S.}~\bibnamefont {Jha}}, \bibinfo {author} {\bibfnamefont {R.~J.}\
  \bibnamefont {Foley}}, \emph {et~al.},\ }\bibfield  {title} {\bibinfo {title}
  {Nearby supernova rates from the lick observatory supernova search--ii. the
  observed luminosity functions and fractions of supernovae in a complete
  sample},\ }\href@noop {} {\bibfield  {journal} {\bibinfo  {journal} {Monthly
  Notices of the Royal Astronomical Society}\ }\textbf {\bibinfo {volume}
  {412}},\ \bibinfo {pages} {1441} (\bibinfo {year} {2011})}\BibitemShut
  {NoStop}%
\bibitem [{\citenamefont {Anderson}\ \emph {et~al.}(2014)\citenamefont
  {Anderson} \emph {et~al.}}]{Anderson:2014hta}%
  \BibitemOpen
  \bibfield  {author} {\bibinfo {author} {\bibfnamefont {J.~P.}\ \bibnamefont
  {Anderson}} \emph {et~al.},\ }\bibfield  {title} {\bibinfo {title}
  {{Characterizing the V-band light-curves of hydrogen-rich type II
  supernovae}},\ }\href {https://doi.org/10.1088/0004-637X/786/1/67} {\bibfield
   {journal} {\bibinfo  {journal} {Astrophys. J.}\ }\textbf {\bibinfo {volume}
  {786}},\ \bibinfo {pages} {67} (\bibinfo {year} {2014})},\ \Eprint
  {https://arxiv.org/abs/1403.7091} {arXiv:1403.7091 [astro-ph.HE]}
  \BibitemShut {NoStop}%
\bibitem [{\citenamefont {{Malmquist}}(1922)}]{1922MeLuF.100....1M}%
  \BibitemOpen
  \bibfield  {author} {\bibinfo {author} {\bibfnamefont {K.~G.}\ \bibnamefont
  {{Malmquist}}},\ }\bibfield  {title} {\bibinfo {title} {{On some relations in
  stellar statistics}},\ }\href@noop {} {\bibfield  {journal} {\bibinfo
  {journal} {Meddelanden fran Lunds Astronomiska Observatorium Serie I}\
  }\textbf {\bibinfo {volume} {100}},\ \bibinfo {pages} {1} (\bibinfo {year}
  {1922})}\BibitemShut {NoStop}%
\bibitem [{esa()}]{esaGaiaMission}%
  \BibitemOpen
  \href {https://www.cosmos.esa.int/web/gaia/science-performance/} {\bibinfo
  {title} {Gaia mission science performance}}\BibitemShut {NoStop}%
\bibitem [{\citenamefont {Reyes}\ and\ \citenamefont
  {Anderson}(2023)}]{Reyes:2022boz}%
  \BibitemOpen
  \bibfield  {author} {\bibinfo {author} {\bibfnamefont {M.~C.}\ \bibnamefont
  {Reyes}}\ and\ \bibinfo {author} {\bibfnamefont {R.~I.}\ \bibnamefont
  {Anderson}},\ }\bibfield  {title} {\bibinfo {title} {{A 0.9\% calibration of
  the Galactic Cepheid luminosity scale based on Gaia DR3 data of open clusters
  and Cepheids}},\ }\href {https://doi.org/10.1051/0004-6361/202244775}
  {\bibfield  {journal} {\bibinfo  {journal} {Astron. Astrophys.}\ }\textbf
  {\bibinfo {volume} {672}},\ \bibinfo {pages} {A85} (\bibinfo {year}
  {2023})},\ \Eprint {https://arxiv.org/abs/2208.09403} {arXiv:2208.09403
  [astro-ph.GA]} \BibitemShut {NoStop}%
\bibitem [{\citenamefont {Vilardell}\ \emph {et~al.}(2010)\citenamefont
  {Vilardell}, \citenamefont {Ribas}, \citenamefont {Jordi}, \citenamefont
  {Fitzpatrick},\ and\ \citenamefont {Guinan}}]{vilardell2010distance}%
  \BibitemOpen
  \bibfield  {author} {\bibinfo {author} {\bibfnamefont {F.}~\bibnamefont
  {Vilardell}}, \bibinfo {author} {\bibfnamefont {I.}~\bibnamefont {Ribas}},
  \bibinfo {author} {\bibfnamefont {C.}~\bibnamefont {Jordi}}, \bibinfo
  {author} {\bibfnamefont {E.~L.}\ \bibnamefont {Fitzpatrick}},\ and\ \bibinfo
  {author} {\bibfnamefont {E.~F.}\ \bibnamefont {Guinan}},\ }\bibfield  {title}
  {\bibinfo {title} {The distance to the andromeda galaxy from eclipsing
  binaries},\ }\href@noop {} {\bibfield  {journal} {\bibinfo  {journal}
  {Astronomy \& Astrophysics}\ }\textbf {\bibinfo {volume} {509}},\ \bibinfo
  {pages} {A70} (\bibinfo {year} {2010})}\BibitemShut {NoStop}%
\bibitem [{\citenamefont {Freedman}\ \emph {et~al.}(2020)\citenamefont
  {Freedman}, \citenamefont {Madore}, \citenamefont {Hoyt}, \citenamefont
  {Jang}, \citenamefont {Beaton}, \citenamefont {Lee}, \citenamefont {Monson},
  \citenamefont {Neeley},\ and\ \citenamefont {Rich}}]{Freedman:2020dne}%
  \BibitemOpen
  \bibfield  {author} {\bibinfo {author} {\bibfnamefont {W.~L.}\ \bibnamefont
  {Freedman}}, \bibinfo {author} {\bibfnamefont {B.~F.}\ \bibnamefont
  {Madore}}, \bibinfo {author} {\bibfnamefont {T.}~\bibnamefont {Hoyt}},
  \bibinfo {author} {\bibfnamefont {I.~S.}\ \bibnamefont {Jang}}, \bibinfo
  {author} {\bibfnamefont {R.}~\bibnamefont {Beaton}}, \bibinfo {author}
  {\bibfnamefont {M.~G.}\ \bibnamefont {Lee}}, \bibinfo {author} {\bibfnamefont
  {A.}~\bibnamefont {Monson}}, \bibinfo {author} {\bibfnamefont
  {J.}~\bibnamefont {Neeley}},\ and\ \bibinfo {author} {\bibfnamefont
  {J.}~\bibnamefont {Rich}},\ }\bibfield  {title} {\bibinfo {title}
  {{Calibration of the Tip of the Red Giant Branch (TRGB)}}\ }\href
  {https://doi.org/10.3847/1538-4357/ab7339} {10.3847/1538-4357/ab7339}
  (\bibinfo {year} {2020}),\ \Eprint {https://arxiv.org/abs/2002.01550}
  {arXiv:2002.01550 [astro-ph.GA]} \BibitemShut {NoStop}%
\bibitem [{\citenamefont {Freedman}\ \emph {et~al.}(2024)\citenamefont
  {Freedman}, \citenamefont {Madore}, \citenamefont {Jang}, \citenamefont
  {Hoyt}, \citenamefont {Lee},\ and\ \citenamefont {Owens}}]{Freedman:2024eph}%
  \BibitemOpen
  \bibfield  {author} {\bibinfo {author} {\bibfnamefont {W.~L.}\ \bibnamefont
  {Freedman}}, \bibinfo {author} {\bibfnamefont {B.~F.}\ \bibnamefont
  {Madore}}, \bibinfo {author} {\bibfnamefont {I.~S.}\ \bibnamefont {Jang}},
  \bibinfo {author} {\bibfnamefont {T.~J.}\ \bibnamefont {Hoyt}}, \bibinfo
  {author} {\bibfnamefont {A.~J.}\ \bibnamefont {Lee}},\ and\ \bibinfo {author}
  {\bibfnamefont {K.~A.}\ \bibnamefont {Owens}},\ }\bibfield  {title} {\bibinfo
  {title} {{Status Report on the Chicago-Carnegie Hubble Program (CCHP):
  Measurement of the Hubble Constant Using the Hubble and James Webb Space
  Telescopes}},\ }\href@noop {} {\  (\bibinfo {year} {2024})}\BibitemShut
  {NoStop}%
\bibitem [{\citenamefont {Riess}\ \emph {et~al.}(2016)\citenamefont {Riess}
  \emph {et~al.}}]{Riess:2016jrr}%
  \BibitemOpen
  \bibfield  {author} {\bibinfo {author} {\bibfnamefont {A.~G.}\ \bibnamefont
  {Riess}} \emph {et~al.},\ }\bibfield  {title} {\bibinfo {title} {{A 2.4 \%
  Determination of the Local Value of the Hubble Constant}},\ }\href@noop {}
  {\bibfield  {journal} {\bibinfo  {journal} {Astrophys. J.}\ }\textbf
  {\bibinfo {volume} {826}},\ \bibinfo {pages} {56} (\bibinfo {year}
  {2016})}\BibitemShut {NoStop}%
\bibitem [{\citenamefont {Scolnic}\ \emph {et~al.}(2018)\citenamefont {Scolnic}
  \emph {et~al.}}]{Pan-STARRS1:2017jku}%
  \BibitemOpen
  \bibfield  {author} {\bibinfo {author} {\bibfnamefont {D.~M.}\ \bibnamefont
  {Scolnic}} \emph {et~al.} (\bibinfo {collaboration} {Pan-STARRS1}),\
  }\bibfield  {title} {\bibinfo {title} {{The Complete Light-curve Sample of
  Spectroscopically Confirmed SNe Ia from Pan-STARRS1 and Cosmological
  Constraints from the Combined Pantheon Sample}},\ }\href
  {https://doi.org/10.3847/1538-4357/aab9bb} {\bibfield  {journal} {\bibinfo
  {journal} {Astrophys. J.}\ }\textbf {\bibinfo {volume} {859}},\ \bibinfo
  {pages} {101} (\bibinfo {year} {2018})},\ \Eprint
  {https://arxiv.org/abs/1710.00845} {arXiv:1710.00845 [astro-ph.CO]}
  \BibitemShut {NoStop}%
\bibitem [{\citenamefont {Marra}\ \emph {et~al.}(2013)\citenamefont {Marra},
  \citenamefont {Amendola}, \citenamefont {Sawicki},\ and\ \citenamefont
  {Valkenburg}}]{Marra:2013rba}%
  \BibitemOpen
  \bibfield  {author} {\bibinfo {author} {\bibfnamefont {V.}~\bibnamefont
  {Marra}}, \bibinfo {author} {\bibfnamefont {L.}~\bibnamefont {Amendola}},
  \bibinfo {author} {\bibfnamefont {I.}~\bibnamefont {Sawicki}},\ and\ \bibinfo
  {author} {\bibfnamefont {W.}~\bibnamefont {Valkenburg}},\ }\bibfield  {title}
  {\bibinfo {title} {{Cosmic variance and the measurement of the local Hubble
  parameter}},\ }\href {https://doi.org/10.1103/PhysRevLett.110.241305}
  {\bibfield  {journal} {\bibinfo  {journal} {Phys. Rev. Lett.}\ }\textbf
  {\bibinfo {volume} {110}},\ \bibinfo {pages} {241305} (\bibinfo {year}
  {2013})},\ \Eprint {https://arxiv.org/abs/1303.3121} {arXiv:1303.3121
  [astro-ph.CO]} \BibitemShut {NoStop}%
\bibitem [{\citenamefont {Odderskov}\ \emph {et~al.}(2017)\citenamefont
  {Odderskov}, \citenamefont {Hannestad},\ and\ \citenamefont
  {Brandbyge}}]{Odderskov:2017ivg}%
  \BibitemOpen
  \bibfield  {author} {\bibinfo {author} {\bibfnamefont {I.}~\bibnamefont
  {Odderskov}}, \bibinfo {author} {\bibfnamefont {S.}~\bibnamefont
  {Hannestad}},\ and\ \bibinfo {author} {\bibfnamefont {J.}~\bibnamefont
  {Brandbyge}},\ }\bibfield  {title} {\bibinfo {title} {{The variance of the
  locally measured Hubble parameter explained with different estimators}},\
  }\href {https://doi.org/10.1088/1475-7516/2017/03/022} {\bibfield  {journal}
  {\bibinfo  {journal} {JCAP}\ }\textbf {\bibinfo {volume} {03}},\ \bibinfo
  {pages} {022}},\ \Eprint {https://arxiv.org/abs/1701.05391} {arXiv:1701.05391
  [astro-ph.CO]} \BibitemShut {NoStop}%
\bibitem [{\citenamefont {Soderberg}\ \emph {et~al.}(2008)\citenamefont
  {Soderberg} \emph {et~al.}}]{Soderberg:2008uh}%
  \BibitemOpen
  \bibfield  {author} {\bibinfo {author} {\bibfnamefont {A.~M.}\ \bibnamefont
  {Soderberg}} \emph {et~al.},\ }\bibfield  {title} {\bibinfo {title} {{An
  Extremely Luminous X-ray Outburst Marking the Birth of a Normal Supernova}},\
  }\href {https://doi.org/10.1038/nature06997} {\bibfield  {journal} {\bibinfo
  {journal} {Nature}\ }\textbf {\bibinfo {volume} {453}},\ \bibinfo {pages}
  {469} (\bibinfo {year} {2008})},\ \Eprint {https://arxiv.org/abs/0802.1712}
  {arXiv:0802.1712 [astro-ph]} \BibitemShut {NoStop}%
\bibitem [{SNe()}]{SNesamegalaxy}%
  \BibitemOpen
  \href {https://www.rochesterastronomy.org/snimages/sndupe.html} {\bibinfo
  {title} {Galaxies with two or more supernovae}}\BibitemShut {NoStop}%
\bibitem [{\citenamefont {Van~Tilburg}\ \emph {et~al.}(2023)\citenamefont
  {Van~Tilburg}, \citenamefont {Baryakhtar}, \citenamefont {Galanis},\ and\
  \citenamefont {Weiner}}]{VanTilburg:2023tkl}%
  \BibitemOpen
  \bibfield  {author} {\bibinfo {author} {\bibfnamefont {K.}~\bibnamefont
  {Van~Tilburg}}, \bibinfo {author} {\bibfnamefont {M.}~\bibnamefont
  {Baryakhtar}}, \bibinfo {author} {\bibfnamefont {M.}~\bibnamefont
  {Galanis}},\ and\ \bibinfo {author} {\bibfnamefont {N.}~\bibnamefont
  {Weiner}},\ }\bibfield  {title} {\bibinfo {title} {{Astrometry with
  Extended-Path Intensity Correlation}},\ }\href@noop {} {\  (\bibinfo {year}
  {2023})},\ \Eprint {https://arxiv.org/abs/2307.03221} {arXiv:2307.03221
  [astro-ph.IM]} \BibitemShut {NoStop}%
\bibitem [{\citenamefont {Galanis}\ \emph {et~al.}(2023)\citenamefont
  {Galanis}, \citenamefont {Van~Tilburg}, \citenamefont {Baryakhtar},\ and\
  \citenamefont {Weiner}}]{Galanis:2023gef}%
  \BibitemOpen
  \bibfield  {author} {\bibinfo {author} {\bibfnamefont {M.}~\bibnamefont
  {Galanis}}, \bibinfo {author} {\bibfnamefont {K.}~\bibnamefont
  {Van~Tilburg}}, \bibinfo {author} {\bibfnamefont {M.}~\bibnamefont
  {Baryakhtar}},\ and\ \bibinfo {author} {\bibfnamefont {N.}~\bibnamefont
  {Weiner}},\ }\bibfield  {title} {\bibinfo {title} {{Extended-Path Intensity
  Correlation: Microarcsecond Astrometry with an Arcsecond Field of View}},\
  }\href@noop {} {\  (\bibinfo {year} {2023})},\ \Eprint
  {https://arxiv.org/abs/2307.06989} {arXiv:2307.06989 [astro-ph.IM]}
  \BibitemShut {NoStop}%
\bibitem [{\citenamefont {{Wang}}\ and\ \citenamefont
  {{Wheeler}}(2008)}]{2008ARA&A..46..433W}%
  \BibitemOpen
  \bibfield  {author} {\bibinfo {author} {\bibfnamefont {L.}~\bibnamefont
  {{Wang}}}\ and\ \bibinfo {author} {\bibfnamefont {J.~C.}\ \bibnamefont
  {{Wheeler}}},\ }\bibfield  {title} {\bibinfo {title} {{Spectropolarimetry of
  supernovae.}},\ }\href
  {https://doi.org/10.1146/annurev.astro.46.060407.145139} {\bibfield
  {journal} {\bibinfo  {journal} {\araa}\ }\textbf {\bibinfo {volume} {46}},\
  \bibinfo {pages} {433} (\bibinfo {year} {2008})},\ \Eprint
  {https://arxiv.org/abs/0811.1054} {arXiv:0811.1054 [astro-ph]} \BibitemShut
  {NoStop}%
\bibitem [{\citenamefont {Bishop}()}]{sn-database-david-bishop}%
  \BibitemOpen
  \bibfield  {author} {\bibinfo {author} {\bibfnamefont {D.}~\bibnamefont
  {Bishop}},\ }\href
  {https://www.physics.purdue.edu/brightsupernovae/snimages/archives.html}
  {\bibinfo {title} {Bright supernovae}}\BibitemShut {NoStop}%
\bibitem [{\citenamefont {Chambers}\ \emph {et~al.}(2016)\citenamefont
  {Chambers}, \citenamefont {Magnier}, \citenamefont {Metcalfe}, \citenamefont
  {Flewelling}, \citenamefont {Huber}, \citenamefont {Waters}, \citenamefont
  {Denneau}, \citenamefont {Draper}, \citenamefont {Farrow}, \citenamefont
  {Finkbeiner} \emph {et~al.}}]{chambers2016pan}%
  \BibitemOpen
  \bibfield  {author} {\bibinfo {author} {\bibfnamefont {K.~C.}\ \bibnamefont
  {Chambers}}, \bibinfo {author} {\bibfnamefont {E.}~\bibnamefont {Magnier}},
  \bibinfo {author} {\bibfnamefont {N.}~\bibnamefont {Metcalfe}}, \bibinfo
  {author} {\bibfnamefont {H.}~\bibnamefont {Flewelling}}, \bibinfo {author}
  {\bibfnamefont {M.}~\bibnamefont {Huber}}, \bibinfo {author} {\bibfnamefont
  {C.}~\bibnamefont {Waters}}, \bibinfo {author} {\bibfnamefont
  {L.}~\bibnamefont {Denneau}}, \bibinfo {author} {\bibfnamefont
  {P.}~\bibnamefont {Draper}}, \bibinfo {author} {\bibfnamefont
  {D.}~\bibnamefont {Farrow}}, \bibinfo {author} {\bibfnamefont
  {D.}~\bibnamefont {Finkbeiner}}, \emph {et~al.},\ }\bibfield  {title}
  {\bibinfo {title} {The pan-starrs1 surveys},\ }\href@noop {} {\bibfield
  {journal} {\bibinfo  {journal} {arXiv preprint arXiv:1612.05560}\ } (\bibinfo
  {year} {2016})}\BibitemShut {NoStop}%
\bibitem [{\citenamefont {Bellm}\ \emph {et~al.}(2018)\citenamefont {Bellm},
  \citenamefont {Kulkarni}, \citenamefont {Graham}, \citenamefont {Dekany},
  \citenamefont {Smith}, \citenamefont {Riddle}, \citenamefont {Masci},
  \citenamefont {Helou}, \citenamefont {Prince}, \citenamefont {Adams} \emph
  {et~al.}}]{bellm2018zwicky}%
  \BibitemOpen
  \bibfield  {author} {\bibinfo {author} {\bibfnamefont {E.~C.}\ \bibnamefont
  {Bellm}}, \bibinfo {author} {\bibfnamefont {S.~R.}\ \bibnamefont {Kulkarni}},
  \bibinfo {author} {\bibfnamefont {M.~J.}\ \bibnamefont {Graham}}, \bibinfo
  {author} {\bibfnamefont {R.}~\bibnamefont {Dekany}}, \bibinfo {author}
  {\bibfnamefont {R.~M.}\ \bibnamefont {Smith}}, \bibinfo {author}
  {\bibfnamefont {R.}~\bibnamefont {Riddle}}, \bibinfo {author} {\bibfnamefont
  {F.~J.}\ \bibnamefont {Masci}}, \bibinfo {author} {\bibfnamefont
  {G.}~\bibnamefont {Helou}}, \bibinfo {author} {\bibfnamefont {T.~A.}\
  \bibnamefont {Prince}}, \bibinfo {author} {\bibfnamefont {S.~M.}\
  \bibnamefont {Adams}}, \emph {et~al.},\ }\bibfield  {title} {\bibinfo {title}
  {The zwicky transient facility: system overview, performance, and first
  results},\ }\href@noop {} {\bibfield  {journal} {\bibinfo  {journal}
  {Publications of the Astronomical Society of the Pacific}\ }\textbf {\bibinfo
  {volume} {131}},\ \bibinfo {pages} {018002} (\bibinfo {year}
  {2018})}\BibitemShut {NoStop}%
\bibitem [{\citenamefont {Tonry}\ \emph {et~al.}(2018)\citenamefont {Tonry},
  \citenamefont {Denneau}, \citenamefont {Heinze}, \citenamefont {Stalder},
  \citenamefont {Smith}, \citenamefont {Smartt}, \citenamefont {Stubbs},
  \citenamefont {Weiland},\ and\ \citenamefont {Rest}}]{tonry2018atlas}%
  \BibitemOpen
  \bibfield  {author} {\bibinfo {author} {\bibfnamefont {J.}~\bibnamefont
  {Tonry}}, \bibinfo {author} {\bibfnamefont {L.}~\bibnamefont {Denneau}},
  \bibinfo {author} {\bibfnamefont {A.}~\bibnamefont {Heinze}}, \bibinfo
  {author} {\bibfnamefont {B.}~\bibnamefont {Stalder}}, \bibinfo {author}
  {\bibfnamefont {K.}~\bibnamefont {Smith}}, \bibinfo {author} {\bibfnamefont
  {S.}~\bibnamefont {Smartt}}, \bibinfo {author} {\bibfnamefont
  {C.}~\bibnamefont {Stubbs}}, \bibinfo {author} {\bibfnamefont
  {H.}~\bibnamefont {Weiland}},\ and\ \bibinfo {author} {\bibfnamefont
  {A.}~\bibnamefont {Rest}},\ }\bibfield  {title} {\bibinfo {title} {Atlas: a
  high-cadence all-sky survey system},\ }\href@noop {} {\bibfield  {journal}
  {\bibinfo  {journal} {Publications of the Astronomical Society of the
  Pacific}\ }\textbf {\bibinfo {volume} {130}},\ \bibinfo {pages} {064505}
  (\bibinfo {year} {2018})}\BibitemShut {NoStop}%
\bibitem [{\citenamefont {Chen}\ \emph {et~al.}(2022)\citenamefont {Chen} \emph
  {et~al.}}]{Chen:2020qnp}%
  \BibitemOpen
  \bibfield  {author} {\bibinfo {author} {\bibfnamefont {P.}~\bibnamefont
  {Chen}} \emph {et~al.},\ }\bibfield  {title} {\bibinfo {title} {{The First
  Data Release of CNIa0.02\textemdash{}A Complete Nearby (Redshift
  \ensuremath{<}0.02) Sample of Type Ia Supernova Light Curves*}},\ }\href
  {https://doi.org/10.3847/1538-4365/ac50b7} {\bibfield  {journal} {\bibinfo
  {journal} {Astrophys. J. Supp.}\ }\textbf {\bibinfo {volume} {259}},\
  \bibinfo {pages} {53} (\bibinfo {year} {2022})},\ \Eprint
  {https://arxiv.org/abs/2011.02461} {arXiv:2011.02461 [astro-ph.HE]}
  \BibitemShut {NoStop}%
\bibitem [{\citenamefont {Tegmark}\ \emph {et~al.}(2006)\citenamefont {Tegmark}
  \emph {et~al.}}]{SDSS:2006lmn}%
  \BibitemOpen
  \bibfield  {author} {\bibinfo {author} {\bibfnamefont {M.}~\bibnamefont
  {Tegmark}} \emph {et~al.} (\bibinfo {collaboration} {SDSS}),\ }\bibfield
  {title} {\bibinfo {title} {{Cosmological Constraints from the SDSS Luminous
  Red Galaxies}},\ }\href {https://doi.org/10.1103/PhysRevD.74.123507}
  {\bibfield  {journal} {\bibinfo  {journal} {Phys. Rev. D}\ }\textbf {\bibinfo
  {volume} {74}},\ \bibinfo {pages} {123507} (\bibinfo {year} {2006})},\
  \Eprint {https://arxiv.org/abs/astro-ph/0608632} {arXiv:astro-ph/0608632}
  \BibitemShut {NoStop}%
\bibitem [{\citenamefont {Anderson}\ \emph {et~al.}(2024)\citenamefont
  {Anderson} \emph {et~al.}}]{Anderson:2024uwn}%
  \BibitemOpen
  \bibfield  {author} {\bibinfo {author} {\bibfnamefont {J.~P.}\ \bibnamefont
  {Anderson}} \emph {et~al.},\ }\bibfield  {title} {\bibinfo {title} {{Optical
  and near-infrared photometry of 94 type II supernovae from the Carnegie
  Supernova Project}},\ }\href {https://doi.org/10.1051/0004-6361/202244401}
  {\bibfield  {journal} {\bibinfo  {journal} {Astron. Astrophys.}\ }\textbf
  {\bibinfo {volume} {692}},\ \bibinfo {pages} {A95} (\bibinfo {year}
  {2024})},\ \Eprint {https://arxiv.org/abs/2410.06738} {arXiv:2410.06738
  [astro-ph.CO]} \BibitemShut {NoStop}%
\bibitem [{\citenamefont {{Pinto}}\ and\ \citenamefont
  {{Eastman}}(2000)}]{2000ApJ...530..744P}%
  \BibitemOpen
  \bibfield  {author} {\bibinfo {author} {\bibfnamefont {P.~A.}\ \bibnamefont
  {{Pinto}}}\ and\ \bibinfo {author} {\bibfnamefont {R.~G.}\ \bibnamefont
  {{Eastman}}},\ }\bibfield  {title} {\bibinfo {title} {{The Physics of Type IA
  Supernova Light Curves. I. Analytic Results and Time Dependence}},\ }\href
  {https://doi.org/10.1086/308376} {\bibfield  {journal} {\bibinfo  {journal}
  {\apj}\ }\textbf {\bibinfo {volume} {530}},\ \bibinfo {pages} {744} (\bibinfo
  {year} {2000})}\BibitemShut {NoStop}%
\end{thebibliography}%

\clearpage

\onecolumngrid
\begin{center}
  \textbf{\large Supplemental Material}\\[.2cm]
  \vspace{0.05in}
  {David Dunsky, I-Kai Chen, Junwu Huang, Ken Van Tilburg, Robert V.~Wagoner}
\end{center}

\twocolumngrid

\setcounter{equation}{0}
\setcounter{figure}{0}
\setcounter{table}{0}
\setcounter{section}{0}
\setcounter{page}{1}
\makeatletter
\renewcommand{\theequation}{S\arabic{equation}}
\renewcommand{\thefigure}{S\arabic{figure}}
\renewcommand{\theHfigure}{S\arabic{figure}}%
\renewcommand{\thetable}{S\arabic{table}}

\onecolumngrid

This Supplemental Material contains supporting treatment on: correcting for observational bias when applying the EEM to a magnitude-limited sample (App.~\ref{sec:dilation}); Lagrange multiplier methods to establish the optimal SN observation strategy for CDL calibration (App.~\ref{sec:optimumSearchStrategyCepheids}) and for a wholly EEM-based $H_0$ inference (App.~\ref{sec:optimumSearchStrategyH0}); 
and the systematic uncertainty in $H_0$ from cosmic variance (App.~\ref{sec:cosmicVariance}).

\section{Supernova Dilatation Bias}
\label{sec:dilation}
We describe a selection bias that can occur when applying the EEM to a magnitude-limited SN sample. Whereas the \emph{intrinsic} asphericity in the LOS direction should equal unity when averaged over a complete sample of SNe in the Universe,  this is not true for a magnitude-limited SNe sample, since the apparent magnitude of a SN depends on both its absolute magnitude, as well as its orientation and shape. This leads to a selection bias.

In ref.~\cite{eem-1}, we studied a parametric model of SNe with a spheroidal shape. The selection bias can be analyzed within the same framework by modeling the photosphere as an isothermal, uniformly emitting spheroid. SNe with identical luminosity can have different \emph{apparent} magnitudes depending on the shape and orientation of the emitting spheroid. For example, an oblate spheroid viewed face-on appears brighter (smaller magnitude) than the same spheroid viewed edge-on. Similarly, a face-on oblate spheroid will generally be brighter (smaller magnitude) than a spherical or prolate spheroid of any orientation (at fixed overall luminosity and distance). 
This dependence of apparent magnitude on shape and orientation introduces a selection bias in magnitude-limited samples: SNe with certain shapes or orientations may be on average more or less distant from the observer, and hence systematically over- or underrepresented in this sample. 
Specifically, in a magnitude-limited SNe sample, the distribution tends to be skewed to larger asphericity, specifically face-on, oblate SNe. This would lead to a bias in the distance measurement if a flat prior is applied to the orientation of the SN, or a prior is placed on the distribution of asphericity based on the measurement of the brightest SN at low redshift. This bias can be estimated and corrected for if a distribution of SN asphericity can be obtained from low-redshift measurements. 

We estimate the bias in the distance measurement as 
\begin{align}\label{eq:biasd}
    \frac{\Delta D_A^\mathrm{bias}}{D_A} 
    \equiv \left\langle \frac{\hat{D}_A-D_A^\mathrm{truth}}{D_A^\mathrm{truth}} \right \rangle \approx 0.1 \sigma_\eta^2 \, ,
\end{align}
where $\eta$ is defined as the ratio between the length along the symmetry axis and the length along the two semi-axes of the SN. To obtain Eq.~\eqref{eq:biasd}, we average over the parameters that determine the shape of the SNe.
The intrinsic distribution of $\eta$ is approximated as a Gaussian distribution with mean $\langle \eta\rangle = 1$ and a small standard deviation $\sigma_\eta$, while the two Euler angles $\vartheta$ and $\varphi$ (see ref.~\cite{eem-1} for definitions) are uniformly distributed over the 2-sphere. A similar scaling for the distance bias of ${\Delta D_A^\mathrm{bias}}/{D_A} \approx 0.1 |\langle \eta\rangle -1|^2$ holds if the intrinsic shape distribution exhibits a slight preference for either prolate or oblate spheroids ($\langle \eta\rangle \neq 1$). For Type IIP SNe, ${\sigma}_{\eta}$ is about $20\%$ based on spectropolarimetric observations~\cite{2008ARA&A..46..433W}, which suggests this bias is below one percent, and can be safely neglected when compared to the other measurement uncertainties of the benchmarks $(\rm{EEM}_{\#1})$ and $(\rm{EEM}_{\#2})$. This selection bias can be sizable when compared to the measurement uncertainties of the benchmarks of futuristic intensity interferometer arrays with $\text{Matchlight}\gtrsim 10^5$. On the other hand, a population-level distribution of the morphological properties of the photosphere and expanding ejecta material can likely be established with high-Matchlight intensity interferometry measurements of \emph{nearby} bright SNe. The statistics of that population could then be used to correct for the selection bias of Eq.~\ref{eq:biasd} in the sample of more distant SNe.
\section{Optimal Observation Strategy: Cepheid Calibration}
\label{sec:optimumSearchStrategyCepheids}

We elaborate on the optimal SN observation strategy to minimize the uncertainty with which the first and second rung of the CDL can be calibrated with the EEM on Type IIP and Ia SNe. 
The optimum is determined using a Lagrange multiplier method.

The Lagrangian to optimize, subject to the constraint that the observation times $t_{{\rm obs},i}$ of each SN add up to $t_\mathrm{obs}$, is
\begin{align}
    L_{\rm CDL} = \sum_{i}\frac{1}{[\frac{\sigma_{D_A}}{D_A}(m_i, t_{{\rm obs},i})]^2 + \sigma_{\rm calibrator}(m_i)^2} - \lambda\left(\sum_{i}t_{{\rm obs},i} - t_\mathrm{obs}\right)  \, ,
\end{align}
where $\sigma_{\rm calibrator} = \sigma_{\rm Cepheid}$ or $\sigma_{\rm Ia}$ depending on whether the EEM is calibrating Cepheids (Eq.~\eqref{eq:anchorVariance}) or directly calibrating Type Ia SNe (Eq.~\eqref{eq:SNIaVariance}). Setting $\partial L_{\rm CDL}/\partial t_{{\rm obs},i} = 0$ and $\partial L_{\rm CDL}/\partial \lambda$ = 0 gives the following set of constraint equations:
\begin{align}
    \label{eq:Lagrangian2SOE}
    \lambda =\frac{\left(\frac{\sigma_{D_A}}{D_A}(m_i,t_0)\right)^2 \frac{t_0}{t_{{\rm obs},i}^2}}{\left([\frac{\sigma_{D_A}}{D_A}(m_i,t_0)]^2 \frac{t_0}{t_{{\rm obs},i}} + \sigma_{\rm calibrator}^2\right)^2};  \qquad \sum_{j}t_{{\rm obs},j} = t_\mathrm{obs} \, .
\end{align}
Here, $t_0 \equiv 60$ hours is the observation time normalization of Eq.~\eqref{eq:sigmaD}. The system of equations in Eq.~\eqref{eq:Lagrangian2SOE} can be solved for $t_{{\rm obs},i}$, which gives the optimal observation of the SN of apparent magnitude $m_i$,
\begin{align}
    \label{eq:optimiumtiCalibrators}
    t_{{\rm obs},i} = \left(t_\mathrm{obs} + \sum_{j}\left[\frac{\frac{\sigma_{D_A}}{D_A}(m_j,t_0)}{\sigma_{\rm calibrator}(m_j)}\right]^2 t_0 \right)\frac{\frac{\sigma_{D_A}}{D_A}(m_i,t_0)/\sigma_{\rm calibrator}(m_i)^2}{\sum_j\frac{\sigma_{D_A}}{D_A}(m_j,t_0)/\sigma_{\rm calibrator}(m_j)^2} - \left(\frac{\frac{\sigma_{D_A}}{D_A}(m_i,t_0)}{\sigma_{\rm calibrator}(m_i)}\right)^2 t_0\, .
\end{align}
Plugging this $t_{{\rm obs},i}$ back into Eq.~\eqref{eq:anchorVariance} or \eqref{eq:SNIaVariance} minimizes the distance modulus uncertainty of the Cepheid or Type Ia SN rung of the cosmic distance ladder.

\begin{figure}
    \centering
    \includegraphics[width=1\textwidth]{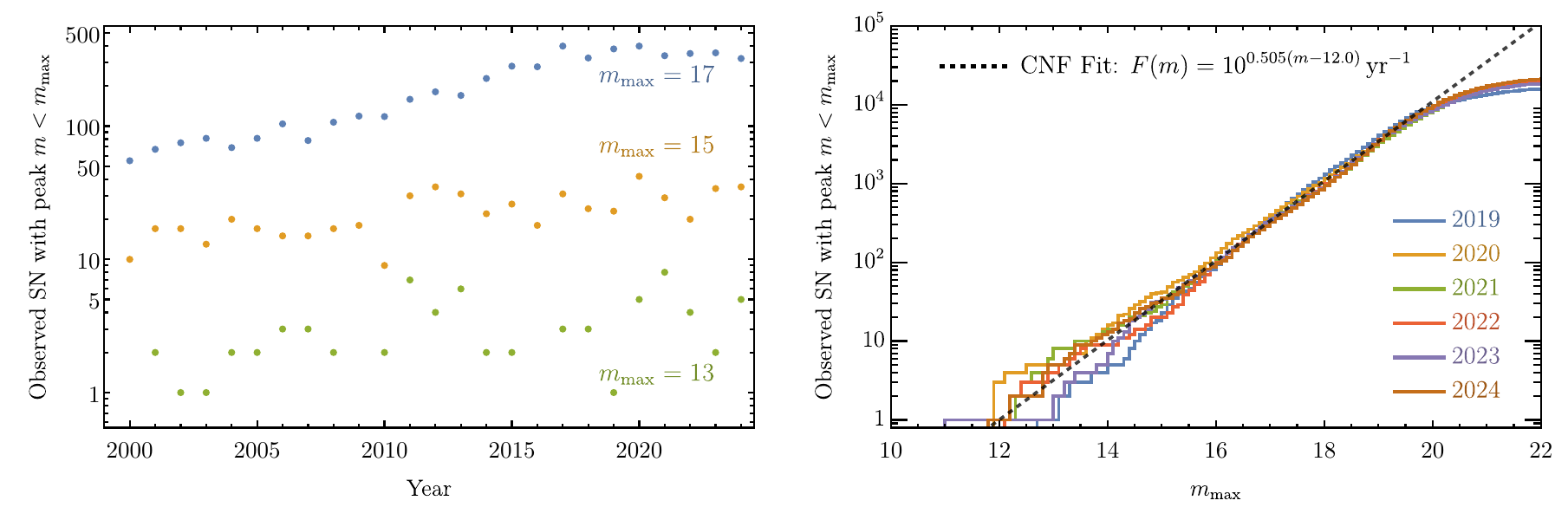}
    \caption{\textit{Left: }Yearly number of observed SNe with peak magnitude $m \leq m_{\rm max}$ over the past 24 years for $m_{\rm max} = 13$ (green), 15 (orange), and 17 (blue). The average number of observed SNe per year has been steady over the past $\sim 20, 15, 8$ years for $m_{\rm max} = 13, 15, 17$, respectively. 
    \textit{Right: }Yearly number of observed SNe with $m \leq m_{\rm max}$ as a function of $m_{\rm max}$ over the past six years. Between $12 \lesssim m_{\rm max} \lesssim 20$, the population is well fit by the cumulative number function (CNF) fit $F(m) = 10^{0.505(m-12.0)} \, {\rm yr}^{-1}$ as shown by the dashed black line. All SNe data taken from the \textit{International Supernovae Network} database \cite{sn-database-david-bishop}.}
    \label{fig:snPopulation}
\end{figure}

To generate realistic SN populations of apparent magnitude $\{m_1,....m_n\}$, we download the list of all SNe observed per year and their apparent magnitudes since $1996$ from ref.~\cite{sn-database-david-bishop}. In recent years, most SNe in this database were detected by the Pan-Starrs \cite{chambers2016pan}, ZTF \cite{bellm2018zwicky},  and ATLAS \cite{tonry2018atlas} surveys. We plot the number of observed SNe per year with $m \lesssim 17$ (15, 13) in the left panel of Fig.~\ref{fig:snPopulation}. 
The right panel of Fig.~\ref{fig:snPopulation} shows the total number of observed SNe with peak magnitude $m < m_{\rm max}$ as a function of $m_{\rm max}$ over the past 6 years, which is well fit by the cumulative number function $F(m) = 10^{0.505(m-12.0)} \, \rm yr^{-1}$, as shown by the dashed gray line in the right panel. A complete sample of SNe will have $F(m)\sim 10^{0.6 m}$, for more discussions about completeness, see \cite{Chen:2020qnp}.

To generate a random population of SNe $\{m_1,....m_n\}$ per year representative of these samples, we extract the PDF $f_{\rm PDF}(m)$ from $F(m)$ by 
\begin{align}
    f_{\rm PDF}(m) \propto F'(m) =  A\times 10^{0.505(m-12.0)} \, {\rm yr^{-1}}  \quad A = \frac{0.505 \ln(10)}{10^{0.505(m_f-12.0)}} \, ,
\end{align}

where $A$ is a normalization factor such that $\int_{-\infty}^{m_f} f_{\rm PDF}(m) dm = 1$ for any arbitrary $m_f > m$. A random annual population of SNe with apparent magnitudes up to $m_f$ is then generated by promoting $m$ to a random variable from the normalized distribution $f_{\rm PDF}$ and calling for $N = f_{\rm SN, type} \times 10^{0.505(m_f-12.0)}$ instances of that random variable (to the nearest integer) via the \texttt{RandomVariate} command of \texttt{Mathematica}, which ensures the cumulative number of SN matches $F(m_f)$. A magnitude-limited survey of SN typically has $f_{\rm SN, Type \,  IIP} \approx 0.17 \times 0.30  \approx 5\%$ and $f_{\rm SN, Type \,  Ia} \approx 79\%$ \cite{li2011nearby}.

If the initial guess $m_f$ is large, the optimal $t_{{\rm obs},i}$ of Eq.~\eqref{eq:optimiumti} for SN with $m$ near $m_f$ will be negative in order for the the total observation time to match $t_\mathrm{obs}$. This scenario is unphysical, as the Lagrange multiplier method does not take into account that $t_{{\rm obs},i} > 0$. In this case, the SN from the generated sample with the largest $m$ is dropped and $t_{{\rm obs},i}$ is redetermined for each SN in this reduced sample. The procedure is repeated until the brightest SN has a $t_{{\rm obs},i} >0$.

Last, the largest $t_{{\rm obs},i}$ in the sample may be greater than $t_{\rm plateau} \approx 6 \times 90$ hours (3 months of nightly observation) for Type IIP SN or $\approx 6\times 30$ hours for Type Ia SN, especially for interferometers with low matchlight which can only measure with good precision the rare, very bright SN in a given year. This scenario is also nonphysical as the Lagrange multiplier method does not know that the SN only lasts for a maximum time of order $t_{\rm plateau}$. In this case, the SN from the generated sample with the largest $t_{{\rm obs},i}> t_{\rm plateau}$ is assigned an observation time of $t_{\rm plateau}$, and the $t_{{\rm obs},j}$ for each SN in this reduced sample is re-determined subject to the new constraint that $\sum_j t_{{\rm obs},j} =t_\mathrm{obs} - t_{\rm plateau}$. The procedure is repeated until the max $t_{{\rm obs},i}$ in the sample is below $t_{\rm plateau}$. 

\section{Cosmic Variance}
\label{sec:cosmicVariance}
\begin{figure}
    \includegraphics[width=1\textwidth]{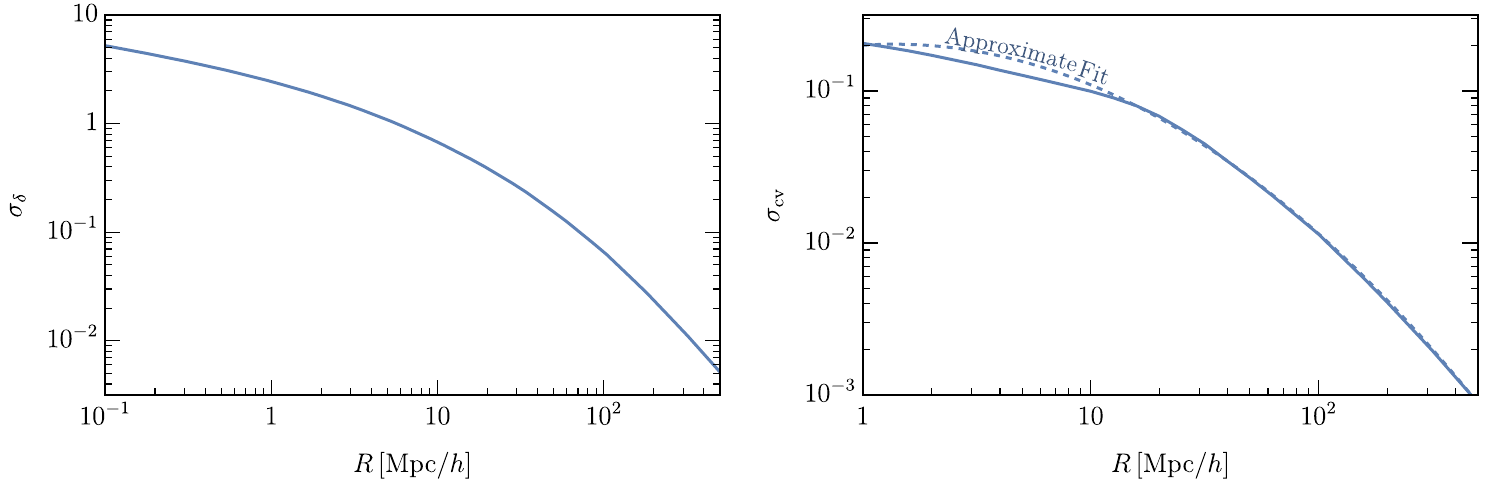}
    \caption{\textit{Left:} Mean standard deviation of a matter density perturbation within a Hubble bubble of radius $R$, assuming the observed matter power spectrum \cite{2020A&A...641A...6P, 2023OJAp....6E..36D, 2022PhRvD.105d3517P,SDSS:2006lmn}. See Eq.~\eqref{eq:meanMatterFluctuation}. \textit{Right:} Mean fluctuation in the expansion rate of a Hubble bubble of radius $R$, assuming the density contrast is a Gaussian random variable with standard deviation $\sigma_\delta$. See Eq.\eqref{eq:sigmaCVDefinition}.}
    \label{fig:cvPlot}
\end{figure}

The inhomogeneous and non-isotropic behavior of the Universe on small scales implies that the Hubble parameter $H$ within a subvolume of the Universe can differ from the global value $H_0$. Within a subvolume possessing an energy density constrast $\delta \rho/\rho$, the local expansion rate will fluctuate relative to the global rate by a relative amount \cite{Marra:2013rba}, 
\begin{align}
    \label{eq:hubbleBubbleFluctuation}
    \frac{\delta H}{H} = -\frac{1}{3} \frac{\delta \rho}{\rho} f(\Omega_m) \Theta\left(\frac{\delta \rho}{\rho}, \Omega_m\right) \, .
\end{align}
Here, $f(\Omega_m) \approx \Omega_m^{0.55}$ and $\Theta(\frac{\delta \rho}{\rho}, \Omega_m) \simeq \Theta(\frac{\delta \rho}{\rho}) \approx 1 - 0.0882\frac{\delta \rho}{\rho} - \frac{0.123 \sin \frac{\delta \rho}{\rho}}{1.29 + \frac{\delta \rho}{\rho}}$ are factors incorporating expansion effects from a non-zero cosmological constant ($\Omega_m \neq 1$) and non-linear density contrasts ($|\delta\rho/\rho| \gtrsim 1$), respectively \cite{Marra:2013rba}. We take $\Omega_m \simeq0.31$ \cite{2020A&A...641A...6P}. Because the chance of a large density fluctuation decreases with larger observational volume, $\delta H/H$ is smaller for more distant SN. Quantitatively, $\sigma_{\rm cv}^2$, which is the systematic variance in Hubble, is the expectation value of the square of Eq.~\eqref{eq:hubbleBubbleFluctuation}, and can be written as
\begin{align}
    \label{eq:sigmaCVDefinition}
    \sigma_{\rm cv}^2(R) = E\left[\left(\frac{\delta H}{H}\right)^2\right] = \int_{X=-1}^{X=\infty} \left[\frac{\delta H}{H}(X)\right]^2 f(X,R)~\dd X\, ,
\end{align}
where $\delta H/H$, Eq.~\eqref{eq:hubbleBubbleFluctuation}, is promoted to a function of the random variable $X = \delta \rho/\rho$ of a Gaussian distribution $f(X,R) = (2\pi \sigma_\delta(R)^2)^{-1/2} \exp[-X^2/2\sigma_\delta^2(R)]$ \cite{Marra:2013rba}. Here, $\sigma_\delta(R)$ is average standard deviation of a density perturbation within a spherical subvolume (a Hubble `bubble') of radius $R$
\begin{align}
    \label{eq:meanMatterFluctuation}
    \sigma_{\delta}^2(R) = \int_{0}^{\infty} \frac{k^2 dk}{2\pi^2} P_m(k)\left[\frac{3 j_1(k R)}{k R} \right]^2 \, ,
\end{align}
where $P_m(k)$ is the matter power spectrum and $j_1$ is the spherical Bessel function of the first kind. The left panel of Fig.~\ref{fig:cvPlot} shows $\sigma_\delta$ as a function of $R$ using the observed matter power spectrum \cite{2020A&A...641A...6P, 2023OJAp....6E..36D, 2022PhRvD.105d3517P,SDSS:2006lmn}. The right panel shows $\sigma_{\rm cv}$ as a function of $R$ after performing the integration on the right-hand-side of Eq.~\eqref{eq:sigmaCVDefinition}. A good approximation for $R \gtrsim 10 \, {\rm Mpc}/h$, is the fitting function 
\begin{align}\label{eq:cvapproximate}
\sigma_{\rm cv,approx} \approx 0.2 \times x^{0.1} 10^{-[\log_{10}x^{0.6}]^2}\qquad x \equiv \left(\frac{R \, h}{\rm Mpc}\right)\, ,
 \end{align}
as shown by the dashed contour of Fig.~\ref{fig:cvPlot}. Here, $h \approx 0.7$ is the dimensionless Hubble constant in units of $100 \, \rm{km/s/Mpc}$. 

\section{Optimal Observation Strategy: Direct Inference of $H_0$}
\label{sec:optimumSearchStrategyH0}

In this section, we elaborate on the optimal SN observation strategy to minimize the statistical uncertainty with which $H_0$ can be inferred (Eq.~\eqref{eq:H0statUncertainty}), as well as systematic uncertainty from cosmic variance (Eq.~\eqref{eq:sigmaCVDefinition}).

\paragraph{Optimal statistical uncertainty} 
First, let us consider the statistical uncertainty for direct measurement of $H_0$, which consists of both the distance uncertainty ${\sigma_{D_A}}/{D_A}$ but also the contribution of peculiar velocity-induced redshift scatter. The statistical uncertainty of observing a single SN with magnitude $m_i$ for a duration of $t_{{\rm obs},i}$ is
\begin{align}
\label{eq:individualHubbleStat}
\left(\frac{\sigma_{H_0,i}}{H_0}\right)_{\rm stat}^2 = \left(\frac{\sigma_{D_A}}{D_A} (m_i, t_{{\rm obs},i})\right)^2 + \frac{\sigma_v^2}{v_i^2}.
\end{align}
where $\frac{\sigma_{D_A}}{D_A}(m_i,t_0) \approx 0.02 \times 10^{0.4(m_i-12.0)} \times (\sigma_t/10\, {\rm ps})^{1/2}(A/\pi(5m)^2)^{-1} (\mathcal{R}/10^4)^{-1/2}(n_{\rm arr}/1)^{-1}(\epsilon/0.5)^{-1}$ as given in Eq.~\eqref{eq:sigmaD} and $\sigma_v \approx 250$ km/s. The velocity $v_i$ follows Hubble's Law $v_i \simeq H_0 d(m_i)$, where for the distance to a SN of apparent magnitude $m_i$, we use luminosity distance $d(m_i) = 10^{(m_i - M_{\rm SN,type} -25)/5}$ Mpc. Note that the precise relation between $v_i$ and $d(m_i)$ is $v_i/H_0 d (m_i) = 1 - \frac{1}{2} (1- q_0) z +\dots $, with the deceleration parameter $q_0 \approx -0.5$ measures the accelerated expansion of the Universe. Practically, since the maximal redshift we can measure SNe with intensity interferometry is $\sim 0.01$, this is a negligible correction. 

\begin{figure}[t]
    \centering
    \includegraphics[width=.497\linewidth]{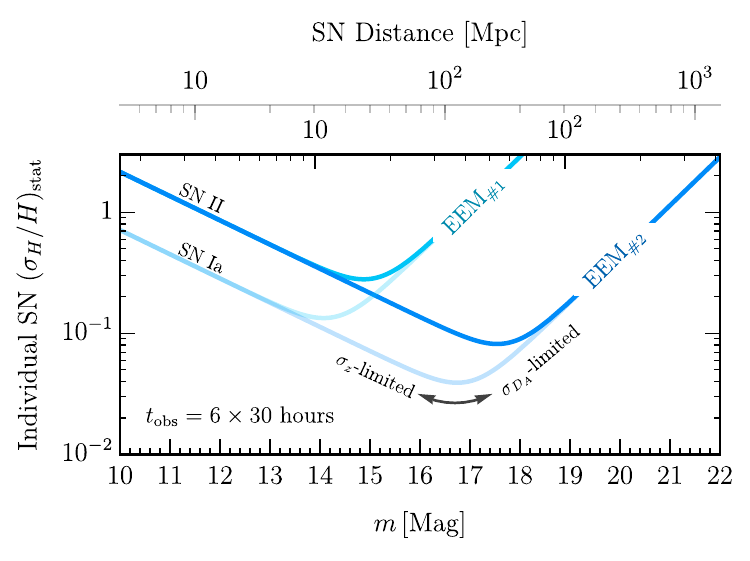}
    \includegraphics[width=.497\linewidth]{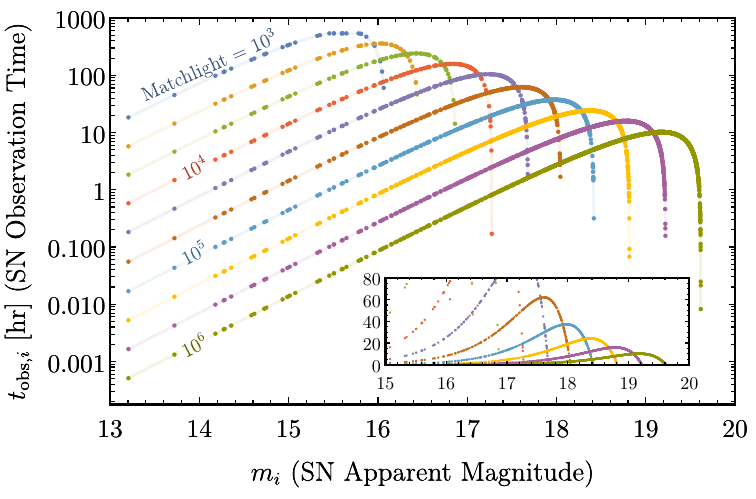}
    \caption{\textit{Left: } EEM measured fractional uncertainty in $H_0$ from observing a \textit{single} SN of apparent magnitude $m$ for $t_{\rm obs} = 6\times 30$ hours. As in Fig.~\ref{fig:systemsDistanceUncertaintyPlot}, the light/medium/dark blue contours are the EEM \#1/2 forecasts on SNe IIP for $\text{Matchlight} = 1{,}250/5 \times 10^4$, respectively; lower-opacity lines are analogous SNe Ia extrapolations. At low apparent magnitudes, the SN velocity scatter is large compared to the Hubble flow velocity ($\sigma_z-$limited region), while at large apparent magnitudes, the uncertainty in the EEM-measured distance is large  ($\sigma_{D_A}-$limited region). The two effects are comparable at the minimum point of the contours. \textit{Right: } An example of a randomly generated SNe population and the corresponding optimal observation allocation times to minimize the statistical uncertainty in $H_0$, Eq.~\eqref{eq:H0statUncertainty}, while adding up to $t_\mathrm{obs} = 5\times365\times6$ hours, according to the Lagrange multiplier description of App.~\ref{sec:optimumSearchStrategyH0}. The colored points show how the same SNe population is allocated differently depending on the Matchlight of the interferometer array performing the EEM measurements. For interferometer arrays of high Matchlight, the peak of the  allocation distribution shifts to larger apparent magnitudes, and is roughly a Gaussian with a standard deviation in apparent magnitude of $\sigma_m \sim 0.46$.}
    \label{fig:sigmaHSingleSN}
\end{figure}

Minimizing ${\sigma_{D_A}}/{D_A}$ favors observing the brightest SNe (smallest magnitude), while minimizing the peculiar velocity, $\sigma_v/v$, favors observing the farthest SNe (largest magnitude) where the peculiar velocity spread is small compared to the Hubble flow velocity. The left panel of Fig.~\ref{fig:sigmaHSingleSN} highlights how these two effects compete with each other. For a fixed observation time $t_{{\rm obs},i}$, there is an optimal magnitude $m_i$ that minimizes the total statistical uncertainty, Eq.~\eqref{eq:individualHubbleStat}, from observing a \emph{single} SN. 

Based on this understanding, we move on to determine the observation strategy to minimize $\hat{\sigma}_{H_0, \rm stat}/H_0$ given a set of SNe with apparent magnitudes $\{m_1, ..., m_n\}$ over 1 year with total observation time $t_\mathrm{obs} = f_{\rm night} \times 1$ yr. The Lagrangian of this setup, subject to the constraint that the sum of the observation times of each SN, $t_{{\rm obs},i}$, add up to $t_\mathrm{obs}$, is
\begin{align}
    L_{H_0,\rm stat} = \sum_{i}\frac{1}{\left[\left(\frac{\sigma_{H_i}}{H_0}\right)_{\rm stat}(m_i, t_{{\rm obs},i})\right]^2} - \lambda\left(\sum_{i}t_{{\rm obs},i} - t_\mathrm{obs}\right)  \, ,
\end{align}
Our goal is to find the $t_{{\rm obs},i}$ that maximizes $L_{H_0, \rm stat}$. This can be done by setting $\partial L_{H_0,\rm stat}/\partial t_{{\rm obs},i} = 0$  and $\partial L_{H_0,\rm stat}/\partial \lambda = 0$, giving the following set of constraint equations:
\begin{align}
    \label{eq:Lagrangian1SOE}
    \lambda =\frac{\left(\frac{\sigma_{D_A}}{D_A}(m_i,t_0)\right)^2 \frac{t_0}{t_{{\rm obs},i}^2}}{\left([\frac{\sigma_{D_A}}{D_A}(m_i,t_0)]^2 \frac{t_0}{t_{{\rm obs},i}} + \frac{\sigma_v^2}{H_0^2 d(m_i)^2}\right)^2} \qquad \sum_{j}t_{{\rm obs},j} = t_\mathrm{obs} \, .
\end{align}
As before, $t_0 \equiv 60$ hours is the observation time normalization of Eq.~\eqref{eq:sigmaD}.

The system of equations Eq.~\eqref{eq:Lagrangian1SOE} can be solved for $t_{{\rm obs},i}$, which gives the optimal observation time of the SN of apparent magnitude $m_i$,
\begin{align}
    \label{eq:optimiumti}
    t_{{\rm obs},i} = \left(t_\mathrm{obs} + \sum_{j}\frac{d(m_j)^2 H_0^2}{\sigma_v^2}\left[\frac{\sigma_{D_A}}{D_A}(m_j,t_0)\right]^2 t_0 \right)\frac{d(m_i)^2 \frac{\sigma_{D_A}}{D_A}(m_i,t_0)}{\sum_j d(m_j)^2 \frac{\sigma_{D_A}}{D_A}(m_j,t_0)} - 
   \frac{d(m_i)^2 H_0^2}{\sigma_v^2}\left[\frac{\sigma_{D_A}}{D_A}(m_i,t_0)\right]^2 t_0 \, .
\end{align}
This optimal time allocation is shown in the right panel of Fig.~\ref{fig:sigmaHSingleSN}, where each point represents a randomly generated SN of apparent magnitude $m_i$. Note that the density of points increases with apparent magnitude $m_i$, making the distribution of time allocation as a function of $m_i$ narrower than it appears in the right panel of Fig.~\ref{fig:sigmaHSingleSN}.
Plugging $t_{{\rm obs},i}$ back into Eq.~\eqref{eq:individualHubbleStat} gives the necessary \textit{individual} fractional uncertainty in Hubble to minimize the \textit{global} statistical fractional uncertainty in Hubble, Eq.~\eqref{eq:H0statUncertainty}.  At large Matchlight, the optimal observational strategy is to target SNe in a narrow range around an optimal apparent magnitude, while at smaller Matchlight, and correspondingly smaller SNe apparent magnitude, the width of the distribution (number of SNe apparent magnitude bins) gets wider. Whereas $\sigma_{D_A}/D_A$ (achieved with SNe in an apparent magnitude bin) only depends on the total amount of time allocated to all the SNe in that apparent magnitude bin, $\sigma_v/v$ depends on the number of SNe in the apparent magnitude bin. Due to the small and finite number of total SNe in an apparent magnitude bin at small apparent magnitude, the distribution has to widen significantly so as to reduce $\sigma_v/v$. When systematic uncertainty is also taken into account, the need for reducing $\sigma_v/v$ also continues to determine the width of the distribution.

\paragraph{Optimal total uncertainty} The optimal observation strategy of allocating $t_{{\rm obs},i}$ to minimize the combination of the statistical $(\hat{\sigma}_{H, \rm stat}/H)$ and systematic $(\hat{\sigma}_{H, \rm sys}/H)$ uncertainties can also be determined via Lagrange multipliers. The statistical uncertainty is the same as in Eq.~\eqref{eq:individualHubbleStat}, while the systematic uncertainty originates from cosmic variance, Eq.~\eqref{eq:sigmaCVDefinition} (see also approximate formula in Eq.~\eqref{eq:cvapproximate}).
The total uncertainty can then be written as
\begin{equation}\label{eq:totalvariance}
    \left(\frac{\sigma_{H_0}}{{H_0}}\right)^2 = \mathbf{w}^T \left(\mathbf{\Sigma}_{\rm stat}+ \mathbf{\Sigma}_{\rm sys}\right)\mathbf{w},
\end{equation}
where $\mathbf{\Sigma}_{\rm stat}$ is a diagonal matrix with  $\left(\mathbf{\Sigma}_{\rm stat}\right)_{ii} = \left(\frac{\sigma_{H_i}}{H_0}\right)_{\rm stat}^2 $ since the statistical uncertainty of different SNe are uncorrelated, while the systematical uncertainty from cosmic variance is $\left(\mathbf{\Sigma}_{\rm sys}\right)_{i,j} = \rho_{ij} \sigma_{\rm cv} (m_i)\sigma_{\rm cv} (m_j)$, since the systematic uncertainty of different SNe are correlated, with correlation parametrized by $\rho_{ij}$. In particular, the cosmic variance uncertainties of SNe with very similar apparent magnitude are almost perfectly correlated, since the parameter $R$ in Eq.~\eqref{eq:cvapproximate} is the distance to a SN $d(m_i)$, depending only on the SN magnitude $m_i$. 

The total uncertainty is a weighted average of the uncertainties from each individual SN with weights $\mathbf{w} = [w_1,w_2,\dots,w_n]^T$, satisfying $\sum_{i} w_i =1$. The optimal weights can be found with the Lagrange multiplier method:
\begin{equation}\label{eq:weights}
    \mathbf{w}_{\rm opt} = \frac{\left(\mathbf{\Sigma}_{\rm stat}+ \mathbf{\Sigma}_{\rm sys}\right)^{-1} \mathbf{1}}{\mathbf{1}^T\left(\mathbf{\Sigma}_{\rm stat}+ \mathbf{\Sigma}_{\rm sys}\right)^{-1} \mathbf{1}},
\end{equation}
where $\mathbf{1} = [1,\dots,1]^T$. The weights in Eq.~\eqref{eq:weights} are then inserted back into Eq.~\eqref{eq:totalvariance} to determine the optimal allocation time $t_{\text{obs},i}$.

Unfortunately, optimizing Eq.~\eqref{eq:totalvariance} does not lead to an analytic, closed-form solution for $t_{{\rm obs},i}$ as in Eq.~\eqref{eq:optimiumti}. As a result, we make the following simplifications:
First, motivated by the observational uncertainties arising from the variation in SN magnitude and line strength evolution after peak light---even for the Type IIP SNe~\cite{1973ApJ...185..303K,Anderson:2024uwn,filippenko1997optical,2000ApJ...530..744P}---and the theoretical understanding that the systematic uncertainty $\sigma_{H_0, \rm sys}$ depends solely on the apparent magnitude of the observed SNe, we bin the SNe into apparent magnitude bins with width $\Delta m = 1/4$ and optimize the total time allocation for the SNe in each apparent magnitude bin instead of optimizing the time allocation of each individual SN, which simplifies the optimization procedure. Second, because the distribution of SNe minimizing $\sigma_{H_0,\rm stat}$ is sharply peaked ($\sigma_m \approx 0.46, \, \text{i.e. FWHM }\simeq 0.46 \times 2 \sqrt{\ln(4)}) = 1.09)$, see Fig.~\ref{fig:sigmaHSingleSN}),
we treat cosmic variance uncertainties as fully correlated, i.~e.~, $\rho_{ij} = 1$, for all SNe in an optimal observing run at fixed Matchlight.
The assumption of a fully correlated cosmic variance permits a simplified analytic formula for the total uncertainty Eq.~\eqref{eq:totalvariance},
\begin{align}
    \label{eq:fullyCorrelatedTotalUncertainty}
     \left(\frac{\sigma_{H_0}}{{H_0}}\right)^2 = \frac{1 + \sum_i N_i\frac{\sigma_{\rm cv}(m_i)^2}{\left(\sigma_{H_0,i}/{H_0}\right)_{\rm stat}^2}}{\sum_i \frac{N_i}{\left(\sigma_{H_0,i}/{H_0}\right)_{\rm stat}^{2}} +\sum_{i,j} N_i N_j\frac{(\sigma_{\rm cv}(m_i)-\sigma_{\rm cv}(m_j))^2}{\left(\sigma_{H_0,i}/{H_0}\right)_{\rm stat}^2 \left(\sigma_{H_0,j}/{H_0}\right)_{\rm stat}^2}} \qquad ( \rho_{ij} = 1  : \text{Fully correlated $\sigma_{\rm cv}$}) \, ,
\end{align}
where $N_i$ is the number of SNe in a given magnitude bin of width $\Delta m$ centered at $m_i$. Note that Eq.~\eqref{eq:fullyCorrelatedTotalUncertainty} is the analytic result for the optimal \textit{weights} $\mathbf{w}_{\rm opt}$, when $\rho_{ij}= 1$; numerical optimization is still required to determine the optimal \textit{observation times} $t_{{\rm obs},i}$. For this, we numerically determine the $t_{{\rm obs,i}}$ that minimize Eq.~\eqref{eq:fullyCorrelatedTotalUncertainty} using the \texttt{NMinimize} command of \texttt{Mathematica} subject to the constraints $\sum_{i} t_{{\rm obs},i} = t_\mathrm{obs}$ and $t_{{\rm obs},i} > 0$.

\begin{figure}
\includegraphics[width=.5\textwidth]{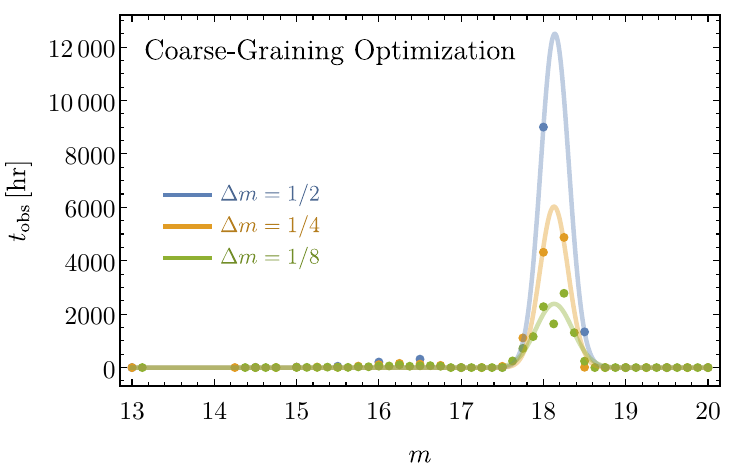}
\caption{
    Example of the coarse-graining procedure to numerically determine the minimum total uncertainty in Hubble, Eq.~\eqref{eq:totalvariance} for fixed Matchlight of $5 \times 10^4$. Each set of colored points is an example of a randomly generate SNe population and the corresponding optimal observation allocation times to minimize the total uncertainty when binning the SNe population into apparent magnitude bins of width $\Delta m = 1/2$ (blue), $1/4$ (orange), and $1/8$ (green). A Gaussian fit overlays the data points to guide the reader and show that allocation time is well described by a Gaussian. As the bin width decreases, the peak of the distribution decreases so that the total observation time remains fixed at $t_\mathrm{obs} = 5 \times 365 \times 6$ hours. 
}
\label{fig:comparisonBindWidthPlot}
\end{figure}

The result of this optimization procedure is shown in Fig.~\ref{fig:comparisonBindWidthPlot} for a series of increasingly finer bin widths at fixed Matchlight $=5\times10^4$. The numerically determined minimum of Eq.~\eqref{eq:fullyCorrelatedTotalUncertainty} is nearly identical across the bin- widths (typical differences $< 1 \%$), signifying rapid convergence for the coarse-graining optimization procedure. In Fig.~\ref{fig:comparisonBindWidthPlot}, the blue, orange, and green dots correspond to SN allocation times in different bins of width $\Delta m = 1/2, 1/4, 1/8$, respectively. For convenience, we overlay a Gaussian fit to the data points to help guide the reader's eye. Note that the peak observation time decreases in height as the bin width decreases so as to conserve total observation time, but the peak FWHM remains essentially unchanged.

Similar to the case with only statistical uncertainty, the optimal time allocation at large Matchlight ($>1{,}000$) is to observe SNe around an optimal apparent magnitude (correspondingly a peak SN observational distance), with the standard deviation of $\sigma_m \approx 0.2$ determined mainly by peculiar velocity uncertainties. The optimal total uncertainty (Eq.~\eqref{eq:totalvariance}) achievable, and the optimal apparent magnitude distribution to observe, does not depend on the choice of bin width $\Delta m$, and the result of $\sigma_m \approx 0.2$ confirms that it is a good approximation to assume the systematic uncertainties to be fully correlated. 
The result of this optimization procedure for a total observation time of $5 \times365 \times 6$ hours in a 5-year span is shown in Fig.~\ref{fig:directH0Plot} in the main text.

\end{document}